\documentclass[]{aa}
\usepackage{natbib,twoopt}
\usepackage{amsmath}
\usepackage{color}
\usepackage[breaklinks=true]{hyperref} 
\bibpunct{(}{)}{;}{a}{}{,}             
\makeatletter
  \newcommandtwoopt{\citeads}[3][][]{\href{http://adsabs.harvard.edu/abs/#3}%
    {\def\hyper@linkstart##1##2{}%
     \let\hyper@linkend\@empty\citealp[#1][#2]{#3}}}
  \newcommandtwoopt{\citepads}[3][][]{\href{http://adsabs.harvard.edu/abs/#3}%
    {\def\hyper@linkstart##1##2{}%
     \let\hyper@linkend\@empty\citep[#1][#2]{#3}}}
  \newcommandtwoopt{\citetads}[3][][]{\href{http://adsabs.harvard.edu/abs/#3}%
    {\def\hyper@linkstart##1##2{}%
     \let\hyper@linkend\@empty\citet[#1][#2]{#3}}}
  \newcommandtwoopt{\citeyearads}[3][][]%
    {\href{http://adsabs.harvard.edu/abs/#3}
    {\def\hyper@linkstart##1##2{}%
     \let\hyper@linkend\@empty\citeyear[#1][#2]{#3}}}
\makeatother

\usepackage{graphicx}
\usepackage{txfonts}
\begin{document}

   \title{Probing star formation and ISM properties using galaxy disk inclination I}
   \subtitle{Evolution in disk opacity since $z\sim0.7$}
   
\authorrunning{S. Leslie et al.}

   \author{S. K. Leslie
          \inst{1}\thanks{Fellow of the International Max Planck Research School 
for Astronomy and Cosmic Physics at the University of Heidelberg (IMPRS-HD)}
          \and M. T. Sargent \inst{2}
          \and E. Schinnerer \inst{1}
          \and B. Groves \inst{3}
          \and A. van der Wel \inst{1}
          \and G. Zamorani \inst{4}
          \and Y. Fudamoto \inst{5}
          \and P. Lang \inst{1}
          \and V. Smol\v{c}i\'{c}\inst{6}
          }

   \institute{Max-Planck-Institut f\"{u}r Astronomie, K\"{o}nigstuhl 17, 69117, 
Heidelberg, Germany
\and Astronomy Centre, Department of Physics and Astronomy, University of 
Sussex, Brighton BN1 9QH, UK
\and Research School of Astronomy and Astrophysics, Australian National 
University, Canberra, ACT 2611, Australia
\and INAF-Osservatorio Astronomico di Bologna, via Gobetti 93/3, 40129, Bologna, 
Italy
\and Observatoire de Gen\'{e}ve, 51 Ch. des Maillettes, 1290 Versoix, 
Switzerland
\and Faculty of Science University of Zagreb Bijeni\v{c}ka c. 32, 10002 Zagreb, 
Croatia\\
\email{leslie@mpia.de}
             }


\abstract{Disk galaxies at intermediate redshift ($z\sim0.7$) have been found in previous work to display more optically 
thick behaviour than their local counterparts in the rest-frame B-band surface 
brightness, suggesting an evolution in dust properties over the past $\sim$6 
Gyr. 
We compare the measured luminosities of face-on and edge-on star-forming 
galaxies at different wavelengths (Ultraviolet (UV), mid-infrared (MIR), far-infrared (FIR), and radio) for two well-matched samples of disk-dominated galaxies: a local Sloan Digital Sky Survey (SDSS)-selected sample at 
$z\sim0.07$ and a sample of disks at $z\sim0.7$ drawn from Cosmic Evolution Survey (COSMOS). We have derived correction factors to account for the inclination dependence of the parameters used for sample selection.
We find that typical galaxies are transparent at MIR 
wavelengths at both redshifts, and that the FIR and radio emission is also transparent as expected. However, reduced sensitivity at these wavelengths limits our analysis; we cannot rule out opacity in the FIR or radio. Ultra-violet  attenuation has increased between $z\sim0$ and $z\sim0.7$, with the 
$z\sim0.7$ sample being a factor of $\sim$3.4 more attenuated. The larger UV attenuation at $z\sim0.7$ can be explained by more clumpy dust around nascent star-forming 
regions.
There is good agreement between the fitted evolution of the normalisation of the SFR$_{\text{UV}}$ versus 1-cos(i) trend (interpreted as the clumpiness fraction) and the molecular gas fraction/dust fraction evolution of galaxies found out to $z<1$.} 



   \maketitle
%

\section{Introduction}

Dust grains in the interstellar medium (ISM) absorb and scatter light emitted by stars and re-emit radiation in the infrared (IR).
Dust often has a large impact on star formation rate (SFR) measurements; for example, global galaxy ultraviolet (UV) fluxes usually suffer from 0-3 magnitudes of extinction \citep{Buat1996}. For global galaxy measurements, the term attenuation is used as it refers to the integrated property of an extended distribution, rather than extinction along a line of sight. Dust attenuation depends on the amount and distribution of dust in a galaxy. 

Galaxy opacity comes mainly from the presence of dust and is therefore highly dependent on the amount of dust, the dust geometry with respect to the stars, the wavelength of light under consideration, and the optical properties of the dust grains. 
Dust exists in spiral galaxies and its distribution can be studied through dust emission in the infrared. Dust has been found out to radial distances 1-2 times the optical radius $R_{25}$ of local galaxies \citep{Smith2016, Hunt2015, DeGeyter2014, Ciesla2012}. Although it also exists on the outskirts of galaxies, the dust distribution has been found to have a lower scale height than the stars \citep{Xilouris1999}. 

Most current models of galaxy dust distributions include at least two dust components: a distribution of diffuse dust and a distribution of optically thick dust clouds at the site of young star-forming regions \citep{DraineLi2007,Galliano2011,Jones2013}. The diffuse dust is heated by a combination of young and old stars and plays a major role in the total
output of dust emission \citep{Ciesla2014,Bendo2012, Boquien2011, Popescu2011, Bendo2010, DraineLi2007, Mathis1990}. Observations have indicated that 
the face-on opacity of the diffuse dust decreases with radius (e.g. 
\citealt{Boissier2004, Popescu2005, Holwerda2005, PerezGonzalez2006, 
Boissier2007,MunozMateos2009}). The diffuse component dominates the dust 
emission at longer wavelengths ($\lambda >160$ $\mu$m) \citep{Dale2007,Draine2007,DeVis}. 
The second important component comes from dust in the birth clouds of massive stars \citep{Charlot2000,daCunha2008}. Dust in birth clouds will experience the strong UV radiation from the young stars, with a radiation field 10-100 times more intense than that experienced by the diffuse component \citep{Popescu2011}. Dust emission from 
these clouds is restricted to the star-forming regions, however it dominates at 
intermediate wavelengths ($\sim$20-60 $\mu$m) \citep{Popescu2008,Popescu2011}. 

In the local Universe, disk galaxies vary from optically thin to thick depending on 
wavelength and location in the disk
\citep{Huizinga1992, Peletier1992, Giovanelli1994, Giovanelli1995, Jones1996,
Moriondo1998, Gonzalez1998}. For example, using the 
background light technique, studies have found that spiral arms are opaque, but 
the opacity of the interarm-region decreases with galactic radius 
\citep{Gonzalez1998,White2000,Keel2001}.

A common method to 
probe the dust opacity, originally proposed by \cite{Holmberg1958, Holmberg1975}, is to study the inclination dependence of starlight (e.g. 
\citealt{Disney1989, Valentijn1990, Jones1996, Tully1998, Maller2009, Masters2010}). 
A disk inclined to our line of sight has higher column density and therefore stellar radiation has to pass through more dust before reaching us, meaning more light is scattered or absorbed. 
Theoretically, for a completely opaque disk, the surface brightness will not change as a function of inclination angle, but for a transparent disk, the surface brightness will increase as the galaxy becomes edge-on. Luminosity trends give complementary information: for an opaque disk, the luminosity will dim with inclination, whereas for a transparent disk the luminosity should be independent of inclination. 
This method requires a statistical sample of galaxies. Recently this method has been used by \cite{Chevallard2013} and \cite{Devour2016} on galaxies in the Sloan Digital Sky Survey (at $z<0.12$) to measure the difference between edge-on 
and face-on attenuation in the optical SDSS ugriz passbands as a function of 
parameters drawn from longer-wavelength surveys that are independent of dust attenuation. 
\cite{Devour2016} found that the strength of the relative attenuation varies strongly 
with both specific star formation rate and galaxy luminosity (or stellar mass), 
peaking at M$_* \approx 3\times 10^{10}$ M$_\odot$.
Our work aims to extend these studies by quantifying the inclination-attenuation relationship for wavelengths commonly used to measure SFR: UV, mid-IR (MIR), 
far-IR (FIR) and radio.

Although studied well locally, the opacity of galaxies as a function of redshift has not been investigated at multiple wavelengths. 
The Cosmic Evolution Survey (COSMOS) covers an area large enough to obtain a significant number of disk galaxies detected at multiple wavelengths. 
Taking advantage of the high-resolution Hubble Space Telescope (HST) imaging of the COSMOS field, \cite{Sargent2010} were able to study the surface brightness-inclination relation of a morphologically well-defined sample of disk-galaxies at $z\sim0.7$.  A direct comparison of COSMOS spiral galaxies with 
artificially redshifted local galaxies revealed that the COSMOS galaxies at 
$0.6<z<0.8$ are on average more opaque than local galaxies, having an almost flat rest-frame B band surface brightness-inclination relation.
\cite{Sargent2010} suggested that this constant relation could be due to the presence of more attenuating material, or a different spatial distribution of dust at $z\sim0.7$. 

In this paper, we extend the analysis of \cite{Sargent2010} to measure the 
inclination dependence of UV, MIR, FIR, and radio luminosity for a sample of 
galaxies at $0.6<z<0.8$ drawn from the COSMOS survey. We compare these results 
to what is found using a local sample, matched in galaxy stellar mass and size, 
drawn from the Sloan Digital Sky Survey (SDSS).

In Section 2 we describe the multi-wavelength data sets used and our approach to select a sample of star-forming disk galaxies. In Section 2.3 we derive and 
apply corrections to measurements of stellar mass, g-band half-light radius, and 
S\'{e}rsic index in order to obtain measurements unbiased by dust-related 
inclination effects. 
In Section 2.4 we show the properties of our sample. We use the inclination-SFR relations to compare the opacity of galaxies in Section 3. In Section 3.3 we fit our UV results with models of attenuation from \cite{Tuffs2004}. In Section 3.4 we show that the UV opacity depends on stellar mass surface density.
In Section 4 we discuss what trends are expected in terms of dust evolution and galaxy opacity at intermediate redshift. 
Our conclusions are presented in Section 5.
We find that galaxies show more overall UV attenuation at $z\sim0.7$, possibly due to a larger fraction of dust that surrounds nascent star-forming regions. 

Throughout this work, we use \cite{Kroupa2001} IMF and assume a flat lambda CDM cosmology 
with ($H_0$, $\Omega_M$, $\Omega_\Lambda$)= (70 km s$^{-1}$ Mpc$^{-1}$, 0.3, 0.7).

\section{Data and sample selection}

To study the inclination dependence of attenuation, we must rely on samples of galaxies that differ only in their viewing angle \citep{Devour2016}.
To achieve a sample of such galaxies both locally and at intermediate redshift 
($z\sim0.7$), we select galaxies from the SDSS and COSMOS survey  regions of the sky, respectively. 

High-resolution imaging is required to accurately fit a model galaxy brightness profile. 
Following \cite{Sargent2010}, we choose to compare a sample at $z\sim0.7$ with a local sample ($z\sim0$), because the central wavelength of the SDSS g band matches the rest-frame wavelengths of objects observed in the F814W filter at redshift $z\sim 0.7$ \citep{Kampczyk2007}. 

At both $z\sim0$ and $z\sim0.7$ we selected star-forming galaxies with stellar 
masses $\log(M_*/\mathrm{M_\odot})>10.2$, g-band half-light radii $r_{1/2}>$5 kpc, 
and S\'{e}rsic index $n<$1.2. These cuts were made to minimise selection biases 
whilst maintaining a reasonable number of galaxies in our $z\sim0.7$ sample and 
will be discussed in more detail in the following sub-sections. 

\subsection{COSMOS $0.6<z<0.8$ disks}
The $z\sim0.7$ COSMOS sample was selected in a similar manner to 
\cite{Sargent2010}, drawn from the complete sample of galaxies with I$\le$22.5 mag 
listed in the Zurich Structure and Morphology Catalog (ZSMC; available on IRSA 
\footnote{https://irsa.ipac.caltech.edu/data/COSMOS/ tables/morphology/}).

We have matched the ZSMC with the COSMOS2015 photometric catalogue of 
\cite{Laigle2016}. The COSMOS2015 catalogue is a near-infrared selected catalogue that uses a combined $\mathrm{z^{++}YJHK_s}$ detection image. 
The catalogue covers a square field of 2 deg$^2$ and uses images from UltraVISTA-DR2, Subaru/Hyper-Suprime-Cam, \textit{Spitzer}, \textit{Herschel}-PACs, and \textit{Herschel}-SPIRE.
Galactic extinction has been computed at the position of each object using the \cite{Schlegel1998} values and the Galactic extinction curve 
\citep{Bolzonella2000,Allen1976}.

We use photometric redshifts (z$_{\text{phot}}$) and stellar masses in the 
COSMOS2015 catalogue \footnote{ftp://ftp.iap.fr/pub/from\_users/ hjmcc/COSMOS2015/}. 
Photometric redshifts were calculated using LePhare 
\citep{Arnouts2002,Ilbert2006} following the method of \cite{Ilbert2013}. 
Fluxes from 30 bands extracted from a 3$''$ aperture using SExtractor were used to calculate the redshift probability distribution. The dispersion of the 
photometric redshifts for star-forming galaxies with $\mathrm{i_{AB}}<23$ is 
$\sigma_{\Delta z/(1+z)}<0.01$. 

We limit our sample to 0.6$\leq z_{\text{phot}}\leq$0.8. At a redshift of 
$\sim$0.7 the observed I-band roughly corresponds to the rest frame g-band 
and the observed NUV-band roughly corresponds to the $z\sim 0$ frame FUV, 
minimising K-corrections and allowing for easy comparison with local samples.

Stellar masses are derived as described in \cite{Ilbert2015}, using a grid of synthetic spectra created using the stellar population synthesis models of 
\cite{Bruzual2003}.
The 90\% stellar mass completeness limit for star-forming galaxies in the redshift of our sample ($0.6<z<0.8$) is 10$^{9}$ 
M$_\odot$ in the UVISTA Deep field \citep{Laigle2016}. We re-scale the stellar 
masses from a Chabrier IMF to a Kroupa IMF by dividing by 0.92 
\citep{Madau2014}.

X-ray detected sources from XMM-COSMOS \citep{Cappelluti2007, Hasinger2007, Brusa2010} are not included in our analysis because of potential active galactic nuclei (AGN) contamination. We have also checked for MIR-AGN using the \cite{Donley2012} IRAC criteria. Galaxies with associations to multiple radio components at 3 GHz are also excluded because they are radio-AGN \citep{Smolcic2017b}. 
We select only galaxies classified as star forming in COSMOS2015 which uses the 
NUV-r/r-J colour-colour selection adapted from the \cite{Williams2009} classification and described in \cite{Ilbert2013}. 

\subsubsection{Morphology information}
Morphological measurements were carried out on HST/Advanced Camera for Surveys 
(ACS) F814W (I-band) images with a resolution of $\sim$0.1\arcsec 
\citep{Koekemoer2007}. This corresponds to a physical 
resolution of $\sim$0.67 and 0.75 kpc at redshifts 0.6 and 0.8, respectively. 
Galaxies are classified as ``early type", ``late type" or ``irregular/peculiar" 
according to the Zurich Estimator of Structural Types algorithm (ZEST; 
\citealt{Scarlata2007}). Late-type galaxies (ZEST Class=2) are separated into 
four sub-classes, ranging from bulge-dominated (2.0) to disc-dominated (2.3) 
systems. 
For simple comparison with the SDSS galaxies, we 
select disk galaxies based on S\'{e}rsic index $n<$1.2 (described in Section 
2.2.1). Increasing the bulge-to-disk ratio at a constant opacity can mimic the effect of increasing the opacity of a pure disk \citep{Tuffs2004}. We have confirmed that our results for the COSMOS sample do not change when selecting pure disks using ZEST Class=2.3 rather than using S\'{e}rsic index. 

Galaxies in the ZSMC catalogue have been modelled with 
single-component S\'{e}rsic (1968) profiles using the \textsc{GIM2D} IRAF 
software package \citep{Simard2002}. In the single-component case, GIM2D seeks 
the best fitting values for the total flux $F_{tot}$, the half-light radius 
$R_{1/2}$, the position angle $\phi$, the central position of the galaxy, the 
residual background level, and the ellipticity $e=1-b/a$, where $a$ and $b$ are 
the semimajor and semiminor axes of the brightness distribution, respectively. See 
\cite{Sargent2007} for more details on how the surface-brightness fitting was 
performed.
In order not to miss low-surface galaxies, we cut our sample at $r_{1/2}>$4 kpc (\citealt{Sargent2007}, see Fig. 12). This is because at a given size a minimum surface brightness is required for a galaxy to meet the I$<$22.5 criterion of the morphological ZSMC catalogue. Including smaller galaxies biases our sample against low brightness galaxies. Our sample is complete for galaxies with $r_{1/2}>4$ kpc. 

\subsubsection{Data used for SFR estimation in the COSMOS field}
The multi-wavelength photometry used for measuring SFRs at FUV, MIR, FIR, and radio wavelengths is summarised in Table \ref{datatab}. For the COSMOS data, all fluxes are found in the COSMOS2015 catalogue,  except for the radio 3GHz fluxes which are from \cite{Smolcic2017b}. 

At $z=0.7$, the GALEX NUV filter roughly corresponds to the FUV filter at $z=0.1$. The COSMOS field was observed as part of the GALEX (Galaxy Evolution Explorer) Deep Imaging Survey (DIS) in the Far-UV ($\sim$1500\AA) and Near-UV  ($\sim$2300\AA).

We also draw 100 $\mu$m fluxes from the COSMOS2015 catalogue.
These data were taken with PACS (Photoconductor Array Camera and Spectrometer; 
\citealt{Poglitsch2010} on the \textit{Herschel} Space Observatory through the PEP (PACS Evolutionary Probe) guaranteed time program \citep{Lutz2011}. Source extraction was performed by a PSF fitting algorithm using the 24 $\mu$m source catalogue to define prior positions.

The VLA-COSMOS 3 GHz Large Project catalogue \citep{Smolcic2017} contains 10830 radio sources down to 5 sigma (and imaged at an angular resolution of 0.74$''$). These sources were matched to COSMOS2015 (as well as other multi-wavelength COSMOS catalogues) by \cite{Smolcic2017b}.

\begin{table*}
\caption{Data used for SFR estimation. The area $\Omega$ quoted is the overlap between the particular wavelength and the SDSS or ACS surveys from which the morphological parameters are drawn. N$_{gals}$ is the number of galaxies in our sample detected at each wavelength with robust mass, inclination, and SFR measurements (Signal-to-noise ratio (S/N)$>$3). }
\label{datatab}
\def\arraystretch{1.2}
\begin{tabular}{|l|l|l|c|c|c|c|}
\hline 
Wavelength & Instrument & Survey & $\Omega$ (deg$^{2}$) & 3$\sigma$ ($\mu$Jy) & N$_\text{gals}$& References\\
\hline
COSMOS UV &GALEX (NUV) &DIS &1.7 & 0.23 & 296&(1)\\
COSMOS MIR &Spitzer MIPS (24$\mu$m)&S-COSMOS& 1.7 &43 & 373 &(2,3)\\
COSMOS FIR &Herschel PACS (100$\mu$m) &PEP&1.7 & 5000 & 87&(4,5)\\
COSMOS Radio &VLA (S-band, 3GHz)&VLA-COSMOS 3GHz& 1.7 & 7 & 58&(6,7)\\
\hline
SDSS UV &GALEX (FUV) &MIS& $\sim$1000 & 1.8 & 1003&(8)\\
SDSS MIR &WISE (W3, 12$\mu$m)&AllWISE& $\sim$8400 &340& 8333&(9,10)\\
SDSS FIR&IRAS (60$\mu$m) & FSC& $\sim$8400 & 1.2$\times$10$^5$&291&(11)\\
SDSS Radio &VLA (L-band, 1.4GHz) &FIRST, NVSS&$\sim$8400 & 450 &198&(12,13,14)\\
\hline

\end{tabular}
\tablebib{ (1) \cite{Capak2007}; (2)\cite{Sanders2007}; (3)\cite{LeFloch2009}; (4)\cite{Poglitsch2010}; (5)\cite{Lutz2011}; (6)\cite{Smolcic2017}; (7)\cite{Smolcic2017b}; (8)\cite{Bianchi2011}; (9)\cite{Wright2010}; (10)\cite{Chang2015}; (11)\cite{Moshir1990}; (12)\cite{Kimball2014}; (13)\cite{Becker1995}; (14)\cite{Condon1998}.}
\end{table*}


\subsection{Local $0.04<z<0.1$ disks in SDSS}
The local sample is drawn from the SDSS Data Release 7 \citep{Abazajian2009}. In particular, we select galaxies with publicly available redshifts, stellar masses, and emission line fluxes from the spectroscopic MPA/JHU 
catalogue\footnote{http://www.mpa-garching.mpg.de/SDSS/DR7/} described in 
\cite{Brinchmann2004}. 
We select galaxies in the redshift range $0.04<z<0.1$. The lower redshift limit is to ensure that the SDSS fibre covers at least 30\% (median coverage 
38\%) of the typical galaxy to minimise aperture effects \citep{Kewley2005}. The sample is constrained to $z<0.1$ to avoid incompleteness and significant evolutionary effects \citep{Kewley2006} and to ensure that the FUV passband lies above the numerous stellar absorption features that occur below 1250\AA.

Total stellar masses were calculated by the MPA/JHU team from ugriz galaxy photometry using the model grids of \cite{Kauffmann2003} assuming a Kroupa IMF. 
The stellar masses in the MPA/JHU catalogue have been found to be consistent with other estimates (see \citealt{Taylor2011, Chang2015}). In the following 
analysis, we use the median of the probability distribution to represent each parameter, and the 16th and 84th percentiles to represent the dispersion.

We further limit our sample to galaxies classified as `starforming' (
$\log(\text{[O\textsc{iii}]/H$\beta$})<0.7-1.2(\log(\text{[N\textsc{ii}]
/H$\alpha$})+0.4)$) or `starburst' (the galaxy is `starforming' but also has an 
H$\alpha$ equivalent width $>$50\AA). 
In this way, we exclude quiescent galaxies and galaxies whose optical emission is dominated by an 
AGN. We also exclude any galaxies that are classified as AGN based on their WISE colours following \cite{Mateos2012}.

\subsubsection{Morphology information}
\cite{Simard2011} provide morphological information that is consistent with the 
data provided by \cite{Sargent2007} used for this paper.
\cite{Simard2011} performed two-dimensional (2D) point-spread-function-convolved 
model fits in the g and r band-passes of galaxies in the SDSS Data 
Release 7. The faint surface brightness limit of the spectroscopic sample was 
set to $\mu_{50}$=23.0 mag arcsec$^{-2}$ for completeness. 
The SDSS images have a typical seeing of 1.4$''$ \citep{Simard2011}. This corresponds to a physical resolution of 1.1 kpc and 2.6 kpc at $z=0.04$ and 0.1, respectively. 
We use the SDSS g-band \textsc{GIM2D} single S\'{e}rsic model fitting results in order to match the rest frame wavelength and model used to derive COSMOS morphological parameters.

To select disk-dominated galaxies, we restrict our sample to galaxies that have a 
g-band S\'{e}rsic index $n<$1.2. 
This is motivated by \cite{Sargent2007}; in particular, their Figure 9, which  
shows that pure disk galaxies (ZEST type 2.3) have in general $n<1.2$.

\subsubsection{Data used for SFR estimation in the SDSS field.}
The catalogues used for measuring SFRs in our local sample are listed in Table \ref{datatab} as well as some information regarding the instruments used, the sky coverage, and average noise properties. 
For all wavelengths, we require a S/N of at least three for the fluxes to be used in our analysis.

We make use of FUV data from the GALEX Medium Imaging Survey (MIS) that has been cross-matched with the SDSS DR7 photometric catalogue by \cite{Bianchi2011}.
We correct the GALEX photometry for galactic reddening following 
\cite{Salim2016} who use the extinction coefficients from \cite{Peek2013} for UV 
bands: \begin{equation}
A_{\text{FUV}} = 10.47 \text{E(B-V)} + 8.59\text{E(B-V)}^2 - 82.2\text{E(B-V)}^3,
\end{equation}
where E(B-V) are the galactic colour-excess values.

We use observations taken by the Wide-field Infrared Survey Explorer (WISE; 
\citealt{Wright2010}) telescope to probe the mid-infrared emission of our local galaxy sample. Specifically, we rely on the catalogue compiled by \cite{Chang2015} which contains a match of SDSS galaxies (in the New York University Value-Added Galaxy Catalog; 
\citealt{Blanton2005,Adelman-McCarthy2008,Padmanabhan2008}) with sources in the AllWISE catalogue. 
Fluxes were measured using the profile-fitting method that assumes the sources are unresolved and has the advantage of minimising source blending.
\cite{Chang2015} found that most galaxies are typically smaller than the PSF, but results from their simulation showed that there is a difference between the true flux and the PSF flux which is a function of the effective radius ($R_e$ of the S\'{e}rsic profile) and this difference is independent of the input flux, axis ratio, or $n$. Therefore, we use their flux values that have been corrected based on $R_e$. 

Long wavelength infrared data were obtained from the IRAS (Infrared Astronomical Satellite) Faint Source Catalog (FSC) v2.0. \citep{Moshir1990}.
We match the FSC to our parent SDSS sample using the \texttt{Sky Ellipses} algorithm in \texttt{Topcat} because the IRAS source position uncertainties are significantly elliptical. We match using the 2$\sigma$ positional uncertainty ellipse of the IRAS sources and assume a circular positional uncertainty on each SDSS source of 1$''$. 

We make use of the radio continuum flux densities compiled in the Unified Radio Catalogue v2 \citep{Kimball2008,Kimball2014}. 
The catalogue was constructed by matching two 20 cm (1.4 GHz) surveys conducted with the Very Large Array (VLA): FIRST (Faint Images of the Radio Sky at Twenty Centimeters; \citealt{Becker1995}) and NVSS (NRAO-VLA Sky Survey; \citealt{Condon1998}). 
The FIRST survey has an angular resolution of $\sim$5\arcsec and a typical rms of 0.15 mJy/beam. The NVSS survey, on the other hand, has an angular resolution of $\sim$45$''$ and a typical rms of 0.45 mJy/beam. We use sources with a $\leq$2$''$ separation between the FIRST and SDSS positions as recommended by \cite{Ivezic2002} and \cite{Kimball2014}. 
For our analysis, we use the 1.4 GHz fluxes from the NVSS catalogue because it is better suited for measuring the global flux from a galaxy than the FIRST survey due to the more compact observing configuration.

\subsection{Inclination-independent sample selection}\label{inccorrections}

\begin{figure*}
\includegraphics[width=0.49\linewidth]{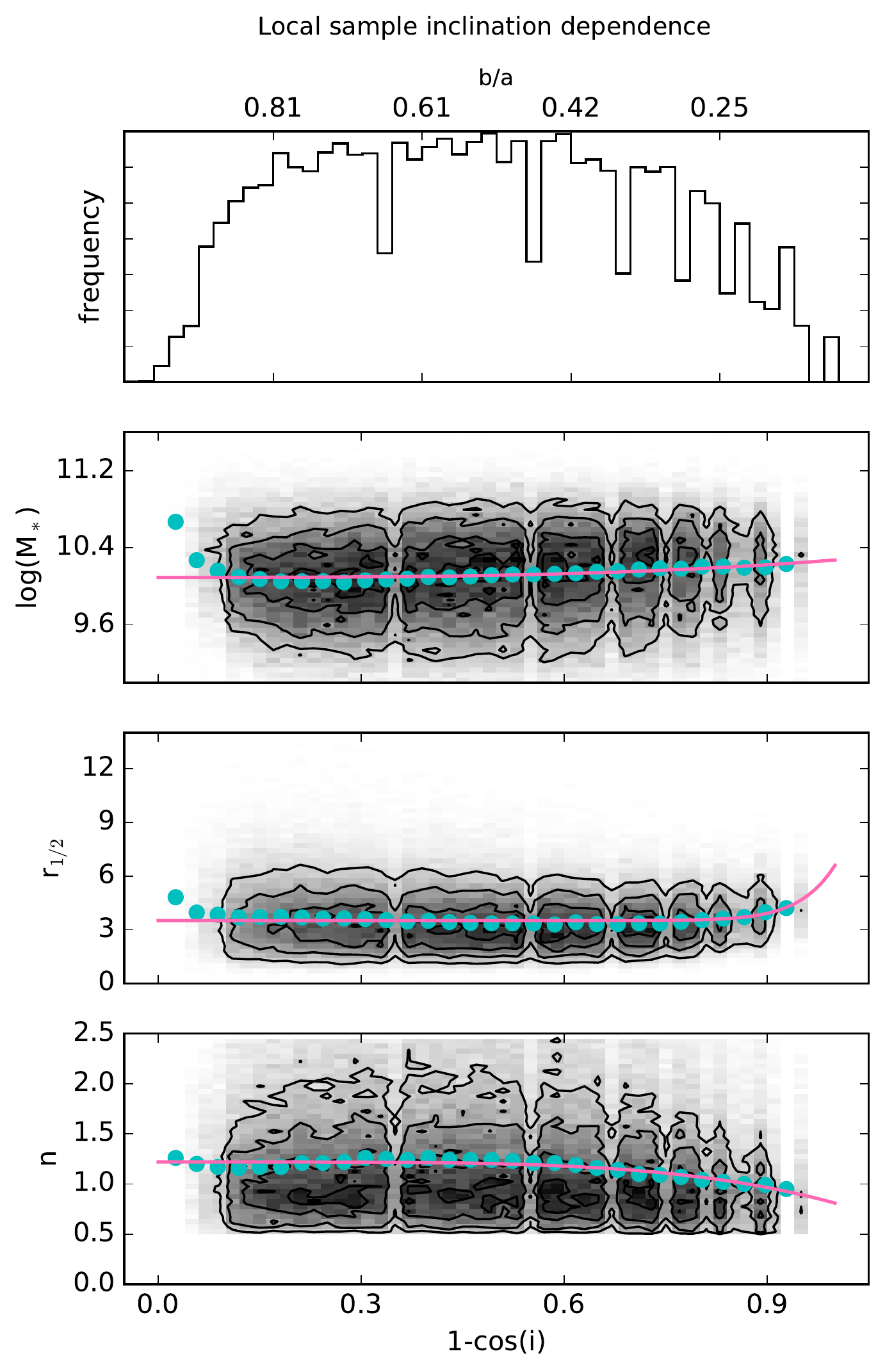}
\includegraphics[width=0.49\linewidth]{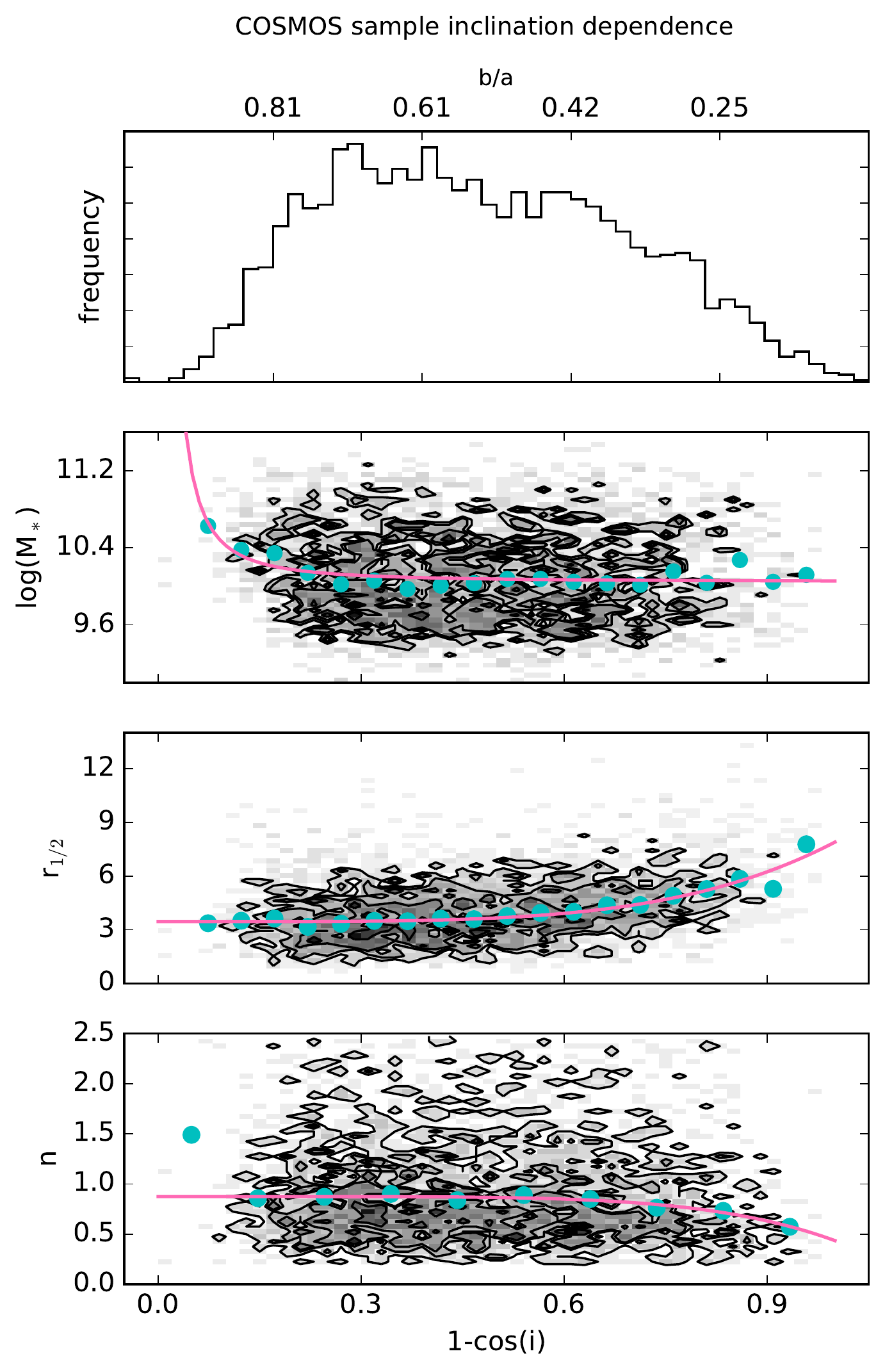}
\caption{Relationship between parameters used for sample selection (stellar mass 
$M_*$ [M$_\odot$], half-light radius in rest-frame g-band $r_{1/2}$ [kpc], and S\'{e}rsic index $n$) 
with inclination (in the form of 1-cos(i)). The distributions are shown for 
galaxies in our parent samples (See Sects. 2.1, 2.2, and 2.3). A histogram of the inclination distribution 
(or equivalently axis ratio distribution) is shown in the top panels. Galaxies 
randomly oriented would produce a flat distribution. Dips in the local $b/a$ distribution are caused by numerical issues in the \cite{Simard2011} fitting.
Teal dots trace the median 
values in bins of inclination and the pink curves show the resulting power-law 
fit (Eq. \ref{pwl}) to the medians. We subtract the inclination dependent term from our parameters before performing our final sample selection to avoid artificially introducing inclination trends in our results.}\label{icorrections}
\end{figure*}

\cite{Devour2016} describe some of the many potential selection effects which could affect the results of an inclination-dependent study such as ours. If any of the parameters used in our selection, such as stellar mass or radius, is dependent on inclination, then we might get a false signal. 
We show the relationship between the parameters used for our sample selection and galaxy inclination in Figure \ref{icorrections}. 

The inclination is calculated from axis ratios measured from the rest-frame 
g-band image using the \cite{Hubble1926} formula
\begin{equation}
\cos^2(i) = \frac{(b/a)^2-(b/a)^{2}_{\text{min}}}{1-(b/a)^{2}_{\text{min}}},
\label{eqinc}\end{equation}
with $(b/a)_\text{min}$ = 0.15 (following \citealt{Sargent2010} and based on 
\citealt{Guthrie1992} and \citealt{Yuan2004}).  If galaxies were randomly oriented disks, we 
would expect to see a flat distribution of 1-$\cos(i)$. This is roughly the case 
for our local parent sample; however, our COSMOS $z\sim0.7$ parent sample has a 
distribution of 1-$\cos(i)$, slightly skewed towards face-on galaxies. 
Instances of $b/a\sim$ 1 and 0 (and hence 1-$\cos(i)$ = 0 and 1) are rare due to 
intrinsic disk ellipticity and the finite thickness of edge-on systems, 
respectively. 
The apparent thickness might vary with the apparent size ($R_{1/2}$), with smaller images appearing rounder, even if images are deconvolved with the PSF 
\citep{Shao2007}. We do not account for this, but it might contribute to the trend we see between inclination and half-light radius in our samples in Figure 
\ref{icorrections}. 

\cite{Driver2007} performed an empirical correction to remove 
inclination-dependent attenuation effects on the turn-over magnitude in B-band 
luminosity function. Inspired by this, we fit a power-law,
\begin{equation}\label{pwl}
y = k_1[1-\cos(i)]^{k_2}+y_c,
\end{equation}
to our selection parameters, where y is the stellar mass, half-light radius, or S\'{e}rsic index. We first 
calculate the median `y' in bins of 1-$\cos(i)$ (50 bins for the local sample 
and 20 bins for the COSMOS $z\sim0.7$ sample) and then find the best fitting 
parameters $k_1$, $k_2$, and $y_c$ using the Python task scipy.optimize.curve\_fit ($\chi^2$ minimisation). For this analysis, we include only galaxies that satisfy our constraints on redshift and star-formation 
activity (see Sects. 2.1 and 2.2), and  $n<4$ (selecting late-type galaxies), which will be referred to as our ``parent samples''.
The fitted parameters are given in Table \ref{corrections}. 

In our SDSS sample, there is a trend that more inclined galaxies (those with a 
low axis ratio) are more likely to have a higher measured stellar mass. On the other hand, in the COSMOS sample, there is no significant ($k_1$=0 within the errors) relation between stellar mass and inclination. The stellar masses for our local sample were derived using optical photometry (and spectroscopic redshifts) only, 
whereas the stellar mass of COSMOS galaxies is derived using 16 photometric bands, including IRAC near-IR data. This might suggest that the mass of the inclined SDSS galaxies is overestimated due to redder optical colours.

We find that the most important inclination effect is that on galaxy size, 
$r_{1/2}$.
Authors such as \cite{Yip2010, Huizinga1992, Mollenhoff2006, Masters2010} and 
\cite{Conroy2010} have all previously reported that galaxy size measurements 
(such as half-light or effective radius) are dependent on galaxy axis ratio b/a, with edge-on galaxies appearing larger at a fixed magnitude. 
Our local results are consistent with results from \cite{Mollenhoff2006} who show that the B-band sizes increase by 10-40\% from face-on to edge-on due to a reduction of the light concentration by centrally concentrated dust in local galaxies. 
Our fit for the dependence of radius on inclination for the COSMOS sample 
results in a size increase of 110\% when $i$ is increased from  0 to 88 degrees. However, the large correction only applies at high inclination, (b/a$<$0.3, $i > 75\deg$).

\cite{Maller2009} and \cite{Patel2012} found that inclination corrections have a strong dependence on S\'{e}rsic index (in the u band and UVJ bands, 
respectively). Having even a small bulge will reduce the ellipticity of the isophotes of a galaxy, resulting in a deficiency of galaxies with $b/a\sim0$. 
 We can see this trend in the spread of inclination values at a given $n$ in 
Figure \ref{corrections}. We also correct for a slight inclination dependency of $n$ with the inclination for the local sample. The distribution of $n$ has a large tail of high $n$-values which cause even our fit to the median $n$ values to pass above the mode of the distribution. 

During our sample selection we account for the inclination dependence using our 
determined fit values (Table 1) and Equation \ref{pwl}. We subtract the inclination-dependent term to correct parameters to their face-on equivalent. To summarise we select galaxies with
\begin{itemize}
\item $ n_{\text{corr}}<1.2$
\item $r_{1/2,\text{corr}}>4$ kpc
\item $\log(\frac{M_*}{\mathrm{M_\odot}})_{\text{corr}}>10.2.$
\end{itemize}

\begin{table}
\caption{Fitted parameters that describe the inclination dependence of galaxy properties that we use for sample selection. We fit the relation $y = 
k_1[1-\cos(i)]^{k_2}+y_c$ to our parent samples of star-forming galaxies based on 
the SDSS and COSMOS surveys.}\label{corrections}
\begin{tabular}{lccc}
\hline
y & $k_1$ & $k_2$ & $y_c$\\
\hline 
SDSS $\log(M_*/\mathrm{M}_\odot)$ & 0.18$\pm$0.04 &  3.0$\pm$1.4 & 10.09$\pm$0.02\\
COSMOS $\log(M_*/\mathrm{M}_\odot)$ & 0.01$\pm$0.01 & -1.6$\pm$0.5& 10.05$\pm$0.04 \\
SDSS $r_{1/2}$ (kpc) & 3.1$\pm$2.5 &  19$\pm$9& 3.52$\pm$0.04\\
COSMOS $r_{1/2}$ (kpc) & 4.4$\pm$0.4 & 4.4$\pm$0.7 & 3.5$\pm$0.1\\
SDSS $n$ & -0.41$\pm$0.05 &  4.5$\pm$0.7 & 1.22$\pm$0.01 \\
COSMOS $n$ & -0.44$\pm$0.05 & 5.6$\pm$1.2 & 0.87$\pm$ 0.01\\
\hline
\end{tabular}
\end{table}

\subsection{Sample properties}

\begin{figure*}
\includegraphics[scale=0.62]{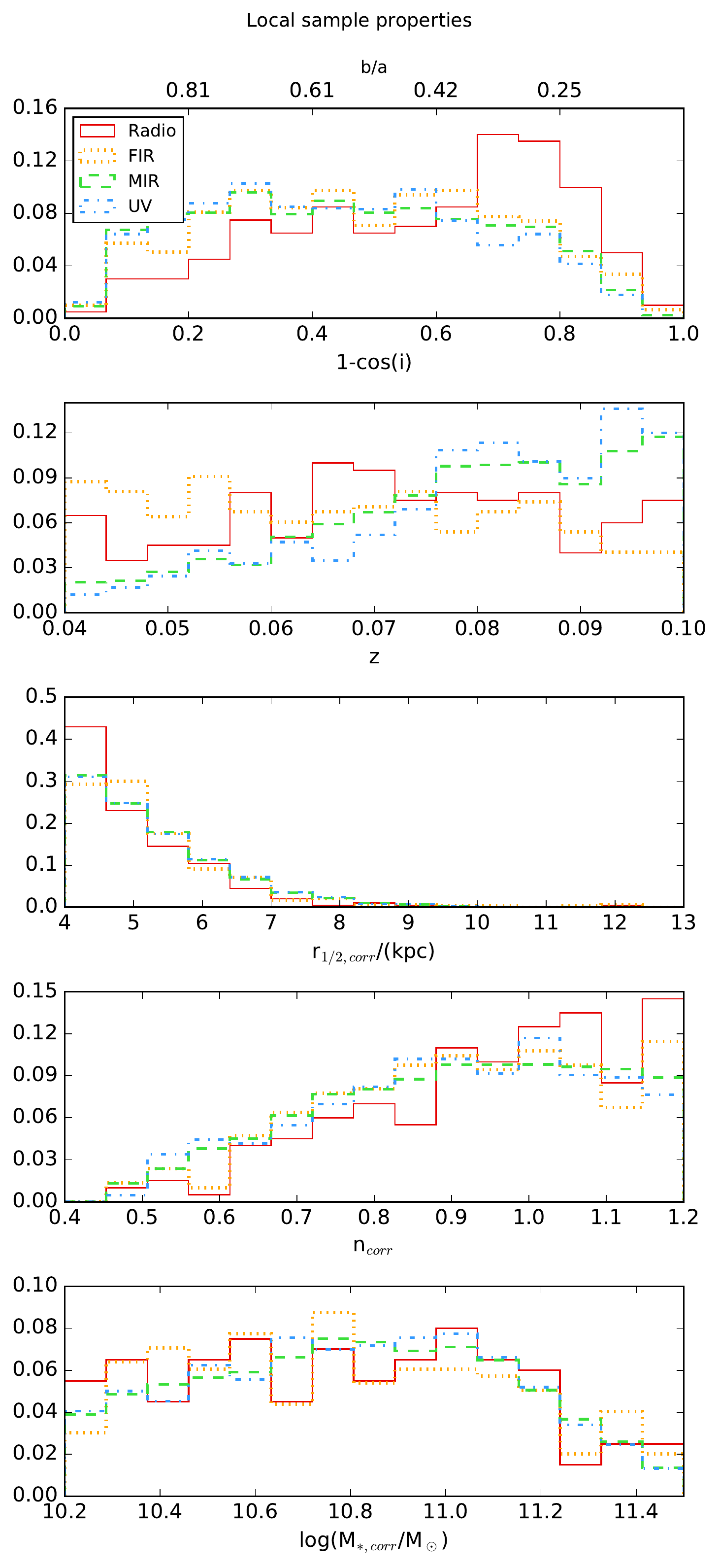}
\includegraphics[scale=0.62]{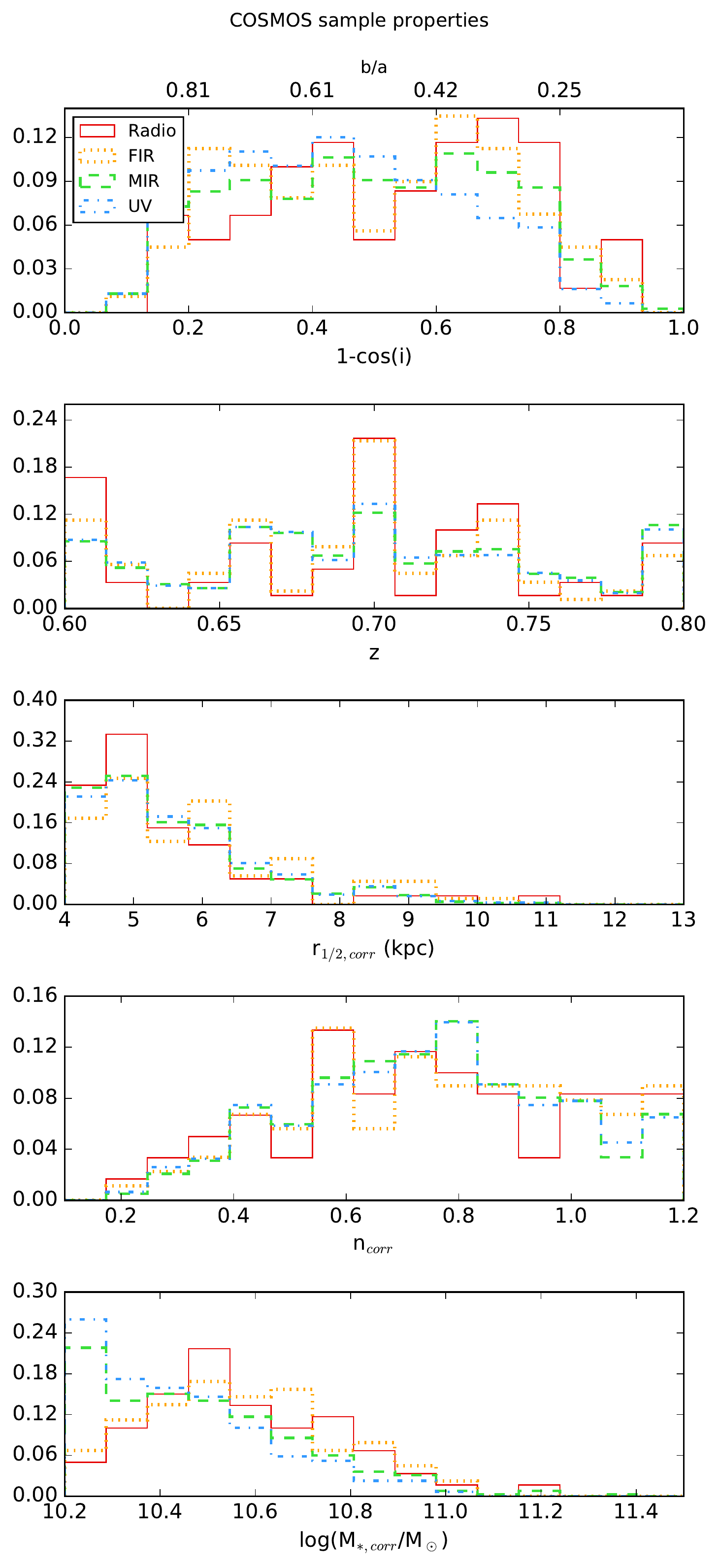}
\caption{Normalised histograms of the galaxy properties of our multi-wavelength 
sub-samples after selection cuts. Histograms show the inclination-corrected values for $r_{1/2}$, $n$ (derived from single 
S\'{e}rsic profile fits to rest-frame g-band light profiles), and log($M_*$). The y-axis shows the fraction of galaxies (in that subsample) in each bin. Radio-detected galaxies have a distribution of optical axis-ratios skewed towards edge-on galaxies. 
}\label{samplep}
\end{figure*}

Figure \ref{samplep} shows the distribution of relevant properties for galaxies in the local sample (left) and the COSMOS $z\sim0.7$ sample (right) after the selection cuts detailed in the previous Section were applied. Each histogram shows the distribution of galaxies detected at a particular wavelength. The total number of galaxies in each subsample is indicated in Table \ref{datatab}. 

The distribution of $\cos(i)$, where $i$ is the inclination, is shown in the top panels of Figure \ref{samplep}. As mentioned above, a sample of randomly oriented thin disks should produce a constant (1-$\cos(i)$) distribution.
In the local sample, galaxies detected in FIR, MIR, and UV catalogues have very 
similar (flat) distributions of galaxy inclination (1-$\cos(i)$).
In both the local and COSMOS sample, the inclination distribution of radio-detected galaxies is skewed towards more inclined galaxies.
In the COSMOS sample, the FIR-detected subsample is also skewed towards more inclined galaxies. 
It is possible that this skewness is caused by a surface-brightness effect -- for a 
fixed global flux limit, an edge-on galaxy will have a higher surface 
brightness than a face-on galaxy because the area on the sky is smaller 
($\propto \sin(i)$). This effect is true especially in the radio, which is unobscured by dust\footnote{We found more galaxies with larger b/a when considering NVSS detected sources compared to FIRST detected sources, but this difference was not statistically significant. }.  

The distributions of redshifts in the COSMOS field are similar for the different wavelengths. Locally, the FIR sample peaks at lower redshifts because of the limited sensitivity of the IRAS catalogue. 
The number of galaxies of the UV and MIR subsamples increases with redshift in 
our local sample as an increasing volume of space is sampled.

The third panels show the half-light radius distributions of the sample after inclination correction. 
The local galaxy sample has a median $r_{1/2,\text{corr}} = 5.04$ kpc, whereas the COSMOS sample has a median of $r_{1/2,\text{corr}} = 5.25$ kpc.
The size distributions of the different subsamples are 
consistent.

The distribution of S\'{e}rsic index, $n$, is shown in the fourth row of Figure 
\ref{samplep}. For our sample, we selected galaxies with $n<1.2$. The 
distribution of $n$ for local galaxies does not go below 0.5 because this was 
the lowest $n$ value allowed by \cite{Simard2011} in their pure S\'{e}rsic 
profile model fits. The distribution of $n$ for local galaxies peaks at higher values than the distribution at $z\sim0.7$. This could be due to star-forming galaxies becoming more mature over time \citep{Scoville2013}. 


The stellar mass distributions (Fig. \ref{samplep}, bottom row) 
show some important biases between the subsamples. For the
COSMOS sample, the radio and FIR subsamples are peaked at larger stellar masses than the UV and MIR subsamples.
In the local sample the FIR 
$\log(M_*)$ distribution starts to fall off at lower stellar masses due to the shallow detection limit of IRAS survey and the correlation between stellar mass and galaxy luminosity. Adopting the relationship between SFR and stellar mass (Eq. \ref{MSeq}), the local FIR observations can only detect a main sequence galaxy with $\log(M_*/\mathrm{M}_\odot)>10.2$ at $z\sim 0.04$. For this calculation, a galaxy with SFR 0.3 dex (1$\sigma$) below the MS given in Section 3 was considered. Similarly, the COSMOS FIR observations are complete down to $\log(M_*/\mathrm{M}_\odot)>10.4$ at $z\sim0.6$.
%

\subsection{SFR estimates}
To calculate SFR, we use the UV and TIR calibrations compiled in \cite{Kennicutt2012}. These calibrations assume a Kroupa IMF and a constant star formation rate. Star formation rates are given in units of M$_\odot$ yr$^{-1}$. 

Young stars emit the bulk of their energy in the rest-frame UV. The FUV emission from star-forming galaxies is dominated by radiation from massive O and B 
stars. 
We use the SFR calibration derived for the GALEX FUV bandpass to calculate 
SFR$_{\text{UV}}$ for our local sample (see Table 1 of \citealt{Kennicutt2012}).
At $z=0.7$ the GALEX NUV filter samples approximately the same rest-frame wavelength as the GALEX FUV filter at $z=0.1$ \citep{Zamojski2007}. The GALEX NUV flux is used with the local FUV calibration. No correction for dust attenuation is applied. FUV attenuation corrections and their inclination dependence will be discussed in Paper II of this series.

For both the mid- and far-IR monochromatic tracers, we convert the monochromatic flux density to the total infrared luminosity (8-1000 $\mu$m) using the
\cite{Wuyts2008} spectral energy distribution (SED) template and then calculate the SFR using the L$_{\text{IR}}$ 
calibration of \cite{Murphy2011}. 
The \cite{Wuyts2008} template was constructed by averaging the logarithm of the \cite{Dale2002} templates resulting in a SED shape reminiscent of M82 \citep{Wuyts2011a}. \cite{Wuyts2008} and \cite{Wuyts2011b} use their template to derive L$_{\text{IR}}$ from monochromatic IR bands and show that the 24$
\mu$m-derived SFR is consistent with a multiband FIR SFR over $0<z<3$. The SED is valid for MS galaxies; however, the use of a single template for all the galaxies may introduce some scatter in the estimate of the FIR luminosity\footnote{We have also confirmed that using the MIR SFR calibrations of \cite{Battisti2015} does not qualitatively change our analysis.}. We do not use templates that rely on a priori knowledge of the SFR and distance from the MS (such as e.g. \citealt{Magdis2012}) because the SFRs could be biased due to inclination effects (see e.g. \citealt{Morselli2016} on how galaxy inclination impacts the measured scatter of the MS).

We calculate the luminosity density $L_{\nu}$, in a given 
filter following \cite{daCunha2008}.
For the IRAS and Spitzer/IRAC photometric systems, the convention is to use the calibration spectrum $\nu C_\nu = \lambda C_\lambda= \text{constant}$ (see \citealt{Beichman1988} and the IRAC data 
Handbook 
\footnote{
http://irsa.ipac.caltech.edu/data/SPITZER/docs/irac/iracinstrumenthandbook/}). 
The Spitzer/MIPS system was calibrated using a black body spectrum of 
temperature 10,000K, such that $C_\lambda = B_\lambda(10000 \text{K})$ (MIPS 
Data 
Handbook\footnote{
http://irsa.ipac.caltech.edu/data/SPITZER/docs/mips/mipsinstrumenthandbook/}).
For the WISE system we assume $C = 1$ \footnote{http://wise2.ipac.caltech.edu/docs/release/allsky/expsup/sec4\_4h.html\#conv2ab}.
For each galaxy, we redshift the SED and convolve it with the appropriate filter response to determine the expected measured luminosity. We then scale the SED to match the measured single-band luminosity and integrate the scaled SED between 8 and 1000$\mu$m 
(rest-frame) to obtain the total infrared luminosity.

Locally we use the corrected WISE 12$\mu$m fluxes from \cite{Chang2015} to calculate SFR$_{\text{MIR}}$. At $z=0.7$, we use MIPS 24$\mu$m fluxes, which correspond to a 
rest-frame 14$\mu$m flux. 
We use IRAS 60$\mu$m luminosity to probe SFR$_\text{FIR}$ of local galaxies, as it probes near the peak of the dust SED.
For galaxies in our COSMOS sample, we use the observed 100 $\mu$m PACS data, which correspond to rest frame $\sim$60 $\mu$m.
Filter responses were obtained from the SVO Filter Profile 
Service\footnote{
http://svo2.cab.inta-csic.es/svo/theory/fps3/index.php?mode=browse}. WISE 12$\mu$m luminosities have been observed to correlate well with H$\alpha$ SFR \citep{Lee2013, Cluver2014} above solar metallicity.

To calculate radio-SFRs in our local SDSS sample, we use empirical 1.4 GHz SFR calibration derived recently by \cite{Davies2017} for star-forming galaxies in the Galaxy and Mass Assembly 
(GAMA;\citealt{Driver2011}) and FIRST surveys. We use the calibration anchored to the UV+TIR SFR (after converting from a Chabrier to Kroupa IMF by adding 0.025 dex; \citealt{Zahid2012}). 
Recently, \cite{Molnar2017} used IR data from \textit{Herschel} and 3GHz VLA observations for a sample of star-forming galaxies in the COSMOS field to measure the infrared-radio correlation out to $z<1.5$. \cite{Molnar2017} separated disk-dominated from spheroid-dominated galaxies using the ZEST catalogue. We use their finding that for disk-dominated galaxies, the IR-radio ratio scales as $q_{\text{TIR}} \propto (1+z)^{-0.04\pm0.01}$, and  by relating the radio luminosity to the TIR luminosity, we are able to use the TIR SFR calibration in \cite{Kennicutt2012}.

For a comparison of the SFR tracers used in this work, we refer the reader to Appendix B.

\section{Inclination-dependent SFR}

 In this Section, we consider the inclination dependence of galaxy SFRs derived from 
multiple wavelengths. There are different galaxies included in the different wavelength subsamples. For the local UV sample, this is largely due to the different survey areas. But different emission mechanisms occurring in different galaxies can result in different SED shapes of galaxies, meaning some galaxies are brighter (and detectable) or dimmer (and non-detectable) at particular wavelengths. We have made a size, mass, and S\'{e}rsic index cut to select star-forming galaxies to ensure that galaxies do not have drastically different SED shapes for both our local and COSMOS samples. Nonetheless, the different depths of the different observations can bias the subsample towards particular galaxies. For example, the radio and FIR subsamples might select more starburst galaxies with high SFRs due to their shallow depth. 

Star-forming galaxies fall on a particular locus in the SFR-stellar mass plane with the logarithmic properties being linearly correlated. We refer to this relation as the star-forming galaxy main sequence (MS; \citealt{Noeske2007}). The normalisation and 
(less strongly) slope of the MS have been reported to evolve with redshift, with galaxies at high redshift having higher SFRs at a given mass than local galaxies. The
normalisation evolution is thought to reflect the increased amount of cold gas available to galaxies at high redshift (e.g. \citealt{Tacconi2013}). 
The average galaxy at a given mass forms more stars at $z\sim0.7$ than the average galaxy at $z\sim0$. To place the SFRs of the two samples on a comparable scale, we normalise our multi-wavelength SFRs by the average MS SFR expected for a galaxy with a given stellar mass and redshift.
The locus of the MS is known to depend on sample selection (e.g. \citealt{Karim2011}) and, in particular, on how actively `star forming' the sample under consideration is.
We adopt the best-fit MS relation for an updated version (M. Sargent, private communication) of the data compilation presented in 
\cite{Sargent2014}.
\begin{equation}\label{MSeq}
\log\left(\frac{\text{SFR}_\mathrm{MS}}{\text{M}_\odot\text{yr}^{-1}}\right) = 0.816 \log\left(\frac{M_*}{\text{M}_\odot}\right) - 8.248 + 3\log(1+z).
\end{equation}


By combining different measurements from the literature that employ different selection criteria, different SF tracers (UV, IR, or radio), and different measurement techniques, \cite{Sargent2014} were able to derive a relation that is representative of the average evolution of the sSFR of main sequence galaxies. The new relation presented in Equation \ref{MSeq} has been recalculated using recent works such as \cite{Chang2015}.
However, the IMF \citep{Chabrier2003} and cosmology (WMAP-7; 
\citealt{Larson2011}) assumed by \cite{Sargent2014}, differs from ours. To convert the relation from a Chabrier to a Kroupa IMF, we add 0.03 dex to the stellar mass \citep{Madau2014, Speagle2014}. The different cosmologies will affect the SFR through the square of the luminosity distance for a given redshift. The correction is redshift dependent: For  $0.04<z<0.1,$ we adjust the 
SFR by -0.004 dex and for $0.6<z<0.8$ we adjust the SFR by +0.006 dex.

Between $z=0.07$ and $z=0.7$, (median redshifts of our two samples) the 
normalisation changes such
that an average $\log(M_*/\mathrm{M}_\odot)=10.5$ main sequence galaxy located exactly at 
the core of the SFR-M$_*$ relation has a SFR of 2.6 M$_\odot$ yr$^{-1}$ at $z=0.07$
and 11 M$_\odot$ yr$^{-1}$ at $z=0.7$. We investigate the affects of using different 
MS relations in Appendix A.

Figure \ref{sfrMS} shows the relation log(SFR$_\lambda$) - log(SFR$_{MS}$) with inclination where 
SFR$_{MS}$ is the main sequence SFR for a galaxy of a given mass. 
To quantify trends present in Figure \ref{sfrMS}, we fit a robust linear model to the data:

\begin{equation}
y=\log(\text{SFR/SFR}_\mathrm{MS}) = \text{slope}\cdot 
\left(1-\cos(i)\right)+\text{intercept}
,\end{equation}
using the statsmodels.api package in Python that calculates an iteratively re-weighted least-squares regression given the robust criterion estimator of \cite{Huber1981}.

To estimate the uncertainties in our results, we resampled the data by drawing (with replacement) a new sample of galaxies 1000 times. We then drew the new
galaxy inclination and SFR/SFR$_{\text{MS}}$ values from the measurement uncertainty distribution. The COSMOS sample has asymmetric errors in inclination and stellar mass, and the local sample has asymmetric errors in stellar mass. To sample these asymmetric distributions, we combined two Gaussian distributions with widths given by the upper and lower limits of the error, normalised so that a new sample data-point is equally likely to lie above or below the measurement. 
Linear regression was performed on each sample drawn, and the best-fitting 
slopes and intercepts with their uncertainties are shown in 
Figure \ref{fit1}. 

Projecting the best-fitting parameter pairs in Figure \ref{fit1} onto the horizontal and vertical axes, we obtain a distribution of the best fitting slopes and intercepts, respectively. The 50th 
(median), 5th and 95th percentiles are given in Table \ref{tab2}. The dashed lines in Figure \ref{sfrMS} indicate the median of the fitted parameters. 

The flux limits of the surveys correspond to a minimum observable luminosity, 
and hence SFR, that increases with redshift. We show the SFR that corresponds to this minimum for our redshift ranges coloured in orange in Figure 
\ref{sfrMS}. The limits show that we are limited in our analysis of the FIR and radio-opacity of galaxies both for the local and $z\sim0.7$ samples, however, they were not taken into account in our fits.

\subsection{Slopes: Opacity}
If disk galaxies are transparent or optically thin
then the luminosities and hence SFRs should not depend on inclination angle. 
In this case, we would see no gradient in Figure \ref{sfrMS}.
If galaxies are optically thick, or opaque, then the luminosity observed can only originate from the stars at the outer surface visible to our line of sight. The projected area 
varies with $\cos(i)$ for a circular disk.
Therefore, for an opaque galaxy, we expect the measured SFR to decrease with the 
inclination angle (and $1-\cos(i)$), producing a negative slope in Figure 
\ref{sfrMS}.

A negative slope is clear in both the local and COSMOS SFR$_{\text{FUV}}$ inclination relations. We find that highly inclined galaxies suffer from more FUV 
attenuation than face-on galaxies, indicating that disk galaxies are optically thick at FUV wavelengths at both redshifts. The fitted values for the slopes are 
-0.79$\pm$0.09 and -0.5$\pm$0.2, for the $z\sim 0$ and $z\sim0.7$ samples.

\cite{Davies1993} explained that when measuring surface-brightness-inclination 
relations to study disk opacity, a survey limited by surface brightness might not 
be able to probe a large enough range of surface brightnesses required to detect a slope, even if some opacity is present.
Similarly, if the survey is not deep enough to probe a range of luminosities, then we might incorrectly conclude that a lack of slope indicates that the disks are transparent. 
We find that the radio and FIR data are not deep enough for us to reliably analyse a SFR-inclination gradient at these wavelengths, due to the potential for selection bias. However, the samples do not show any inclination trend in the upper envelope SFR (not affected by limits) probed by the surveys. Secondly, the fraction of limits to detections is approximately constant in bins of inclination.  Therefore, our data are consistent with galaxies being transparent at FIR and radio wavelengths.

The FUV samples give slightly shallower slopes at $z\sim0.7$ than at $z\sim 0$. This could be partially influenced by the flux limits of the COSMOS surveys meaning that only galaxies with relatively large SFR/SFR$_{MS}$ are detected. These limitations were not taken into account in our linear model. 


The $z\sim0$ and $z\sim0.7$ slopes for the UV are consistent at the 1.65 sigma level. 
In Appendix A we show that using the SFR$_{12\mu\text{m}}$ for normalisation rather than the SFR$_\text{MS}$ gives fully consistent slopes of -0.6 at both redshifts. This suggests that no significant evolution has occurred in the dust properties or dust-star geometry between $z=0$ and $z=0.7$.

\subsection{Intercept: SFR calibration and selection effects}
The y-intercept represents how the SFR measured for a face-on galaxy compares to the main sequence SFR at our given redshifts and stellar mass. 
In absence of other effects such as foreground dust at high scale height,
the y-intercept approximately represents the ratio between SFR$_{\lambda}$ of a 
face-on galaxy and the average total SFR (i.e. SFR$_{\text{IR}}$+SFR$_{\text{UV}}$) of a MS 
galaxy.

The UV SFR observed is lower than the MS SFR at all given inclinations, 
resulting in a negative y-intercept. 
We find that the UV SFR under-predicts the MS SFR by $\sim$0.74 dex at $z\sim0.7$, 
compared to $\sim$0.2 at $z\sim 0$. This implies that a larger fraction of the stellar light is attenuated and re-emitted in the infrared in the higher-redshift galaxies. We discuss this 
further in Sections \ref{tuffssection} and  \ref{discussion}. 

The distributions of y-intercepts for the FIR and radio wavelength subsamples, all of them positive,
are dominated by sample-selection effects, in particular, the flux limit of the surveys. 
The radio and infrared samples lie above the MS due to their shallow observations not being able to detect emission from MS galaxies. 

If the SFR calibrations or main-sequence relations were incorrect, 
we would expect this to alter the absolute offset value.  However, using a `concordance' MS evolution fitted to a compilation of data (as done here) is expected to average out differences in normalisation arising due to different methods or SFR tracers. Normalising the SFR$_{\text{UV}}$ to the SFR$_\text{MIR}$ for galaxies detected in both UV and MIR wavelengths gave intercept values of  -0.27$\pm$0.03 and -1.0$\pm0.1$ for the $z\sim 0$ and $z\sim0.7$ samples (App. A.2). However, this result is still dependent on the accuracy of the SFR calibrations at each redshift (see App. B). 

\begin{figure*}[h]
\includegraphics[scale=0.62]{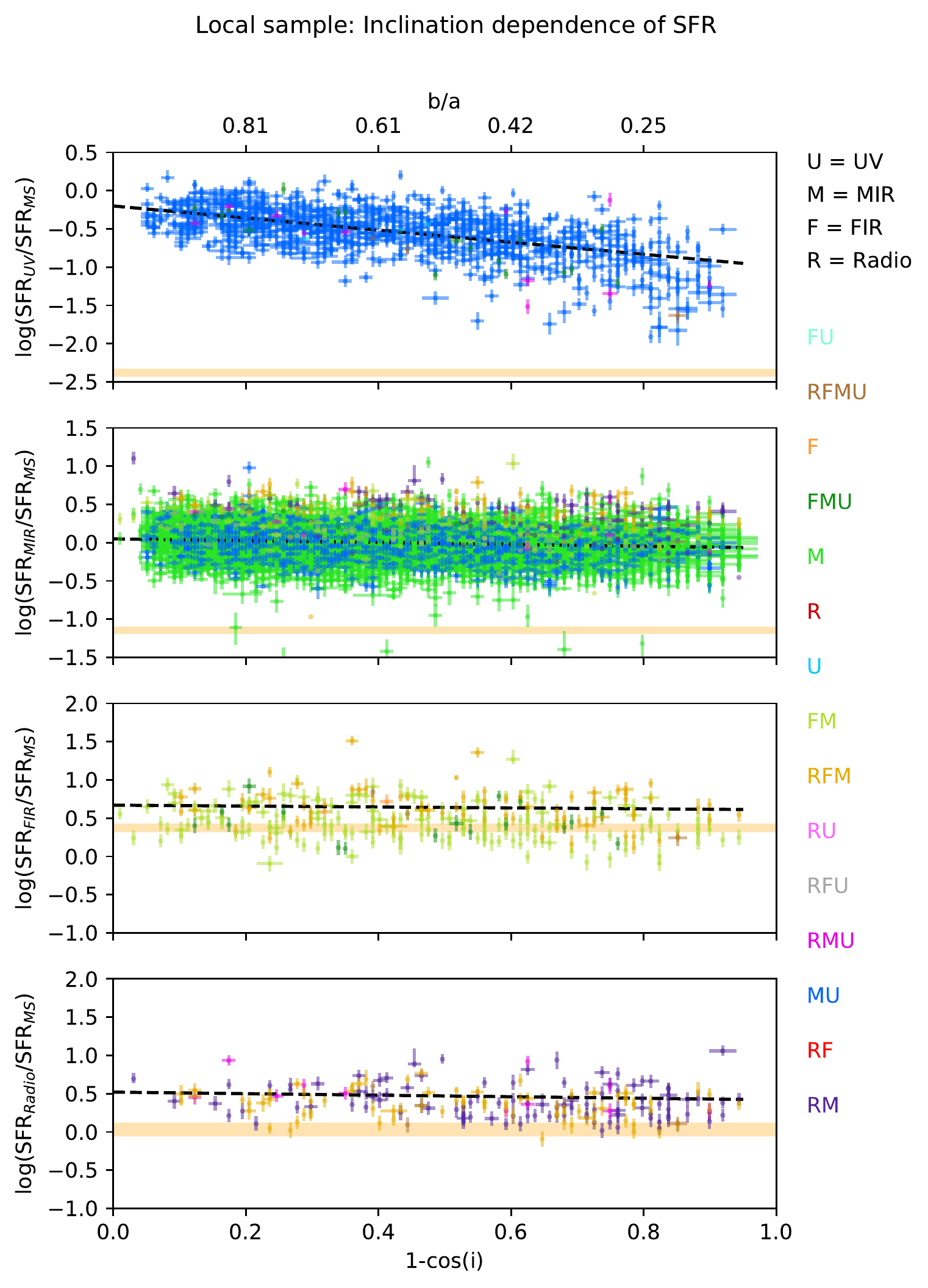}
\includegraphics[scale=0.62]{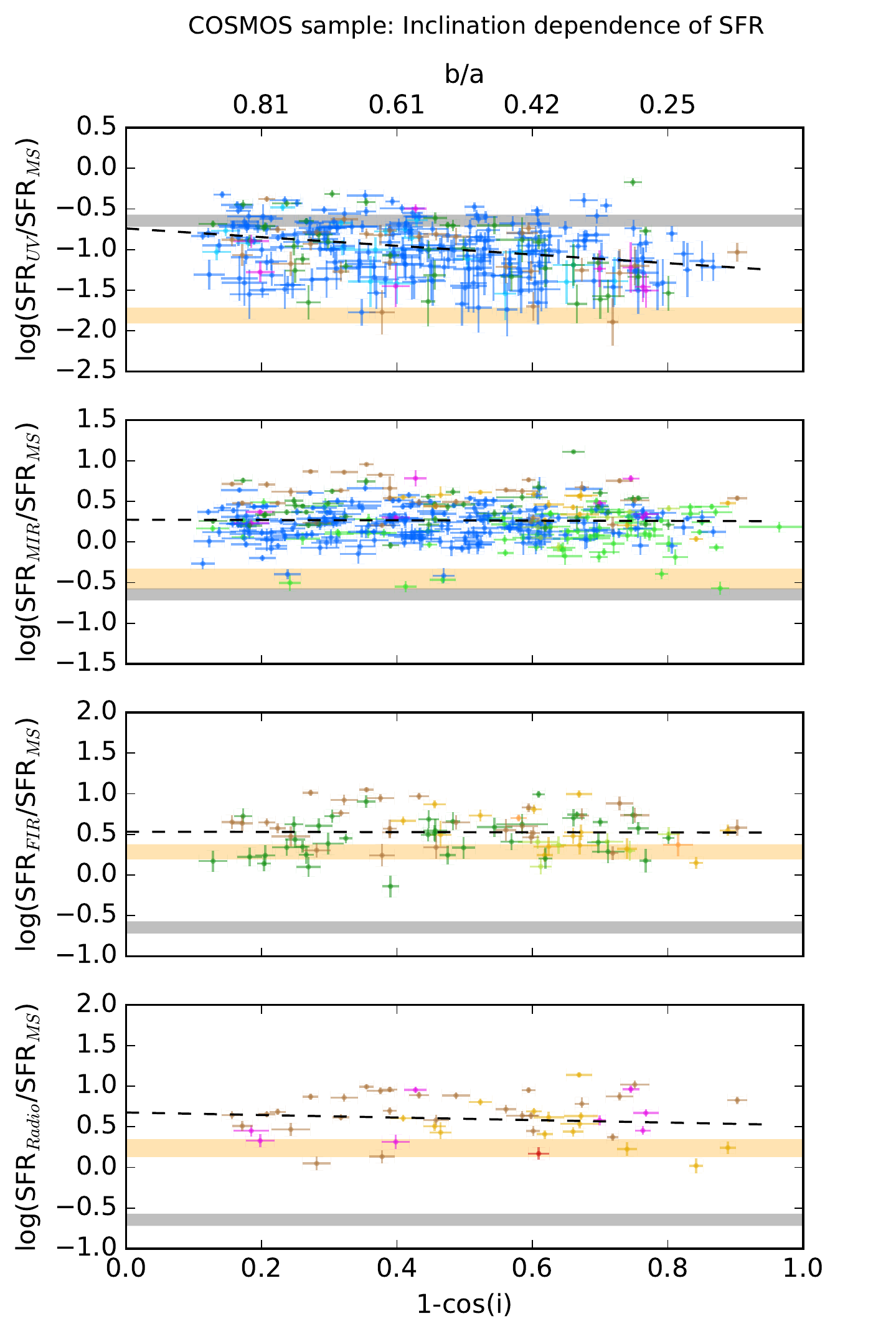}
\caption{Inclination dependence of multiwavelength SFRs. y-axis: log(SFR$_{\lambda}$) - log(SFR$_{\text{MS}}$). Top to bottom shows UV, 
MIR, FIR, and radio-determined SFR. x-axis: 1-cos($i$), where 
$i$ is the galaxy inclination. Face-on galaxies are on the left of the plot (at 
0) and edge-on galaxies are on the right-hand side (at 1). The left panels show 
SDSS galaxies ($0.04<z<0.1$) and the right panels show COSMOS galaxies 
($0.6<z<0.8$). The colour index at the centre indicates which bands were detected; R=radio, F=FIR, M=MIR, and U=UV). The orange shaded region shows the minimum SFR that could be detected across the redshift ranges given the flux limits of the respective surveys. The dashed lines show the median best-fit 
obtained from 1000 realisations of the data.
The grey regions on the right panels show where the local ($0.04<z<0.1$) main-sequence SFR lies with respect to the main sequence at $z=0.7$. This shows how the normalisation of the main sequence evolves over our 
redshift range.}\label{sfrMS}
\end{figure*}

\begin{figure}
\includegraphics[width=\linewidth]{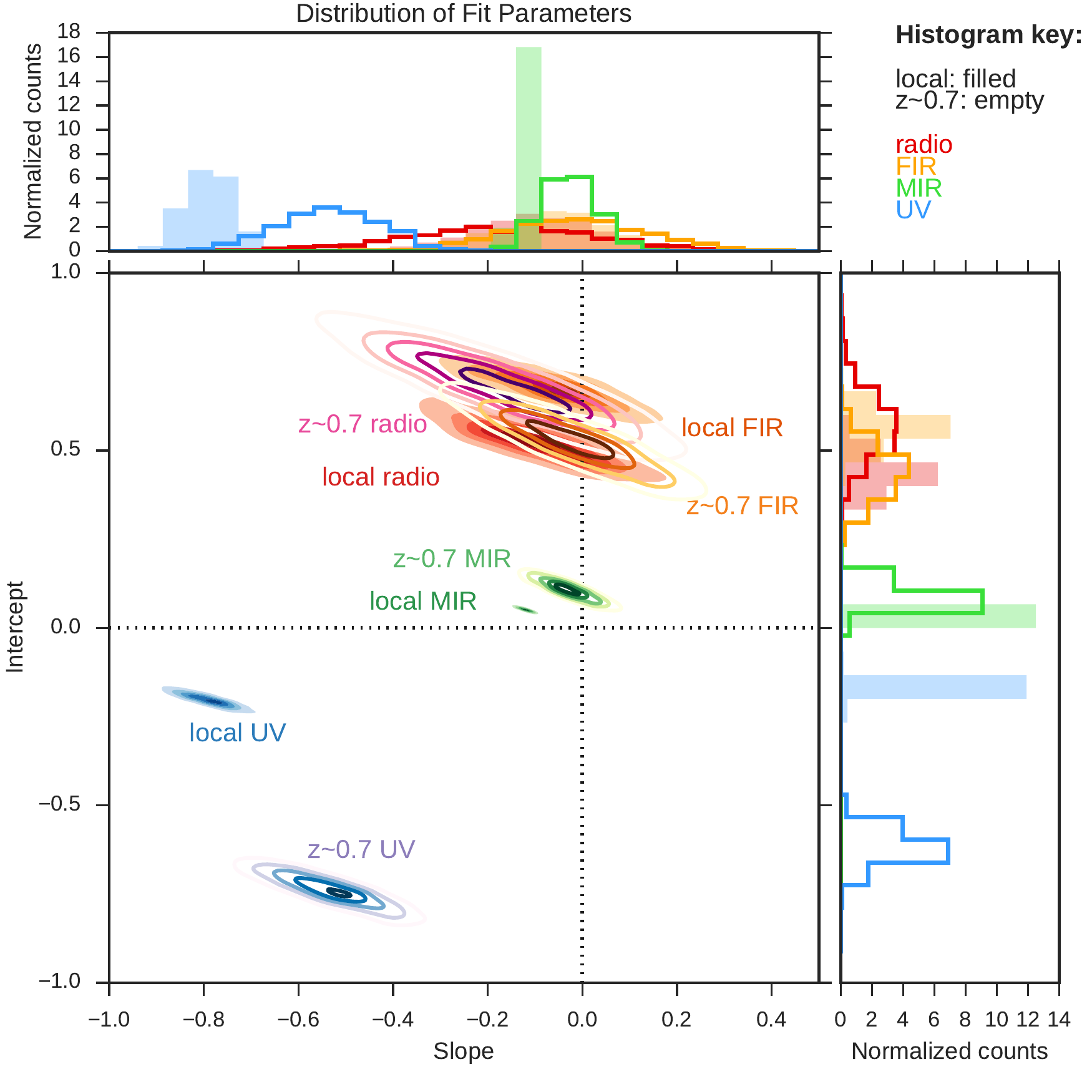}
\caption{Joint and marginalised distributions of the linear model parameters fit 
to the data shown in Figure 3. Filled histograms correspond to SDSS-based samples and the open histograms correspond to COSMOS-based samples. Distributions for the slope and the intercept are 
shown with histograms normalised so that they encompass an area of 1. The median 
of each parameter distribution is given in Table 3. 
}\label{fit1}
\end{figure}

The MS has a scatter of $\sim$0.3 dex \citep{Schreiber2015}, 
meaning that some of the scatter seen in Figure \ref{sfrMS} is due to the fact that galaxies might have true SFRs higher or lower than the average MS SFR used for normalisation. 

\begin{table}
\caption{Best-fit parameters of SFR vs. inclination where SFR is normalised by 
the MS SFR given the galaxy mass: Median, 5th and 95th percentile 
of the distribution.}
\def\arraystretch{1.5}
\centering
\begin{tabular}{|c|cc|}
\hline
$\lambda$ & Slope & Intercept\\
\hline
SFR$_{  UV}$    $z\sim 0$ &     -0.79   $^{+    0.08    }_{-    0.09    }$ &       -0.20   $^{+    0.04    }_{-    0.03    }$ \\ 
SFR$_{  MIR}$   $z\sim 0$ &     -0.12   $^{+    0.03    }_{-    0.02    }$ &       0.05    $^{+    0.01    }_{-    0.01    }$ \\
SFR$_{  FIR}$   $z\sim 0$ &     -0.06   $^{+    0.22    }_{-    0.20    }$ &       0.67    $^{+    0.09    }_{-    0.09    }$ \\
SFR$_{  radio}$ $z\sim 0$ &     -0.10   $^{+    0.20    }_{-    0.22    }$ &       0.52    $^{+    0.11    }_{-    0.09    }$ \\
\hline
SFR$_{  UV}$    $z\sim0.7$ &    -0.54   $^{+    0.18    }_{-    0.17    }$ &       -0.74   $^{+    0.08    }_{-    0.08    }$ \\ 
SFR$_{  MIR}$   $z\sim0.7$ &    -0.03   $^{+    0.10    }_{-    0.09    }$ &       0.28    $^{+    0.05    }_{-    0.05    }$ \\
SFR$_{  FIR}$   $z\sim0.7$ &    -0.02   $^{+    0.23    }_{-    0.25    }$ &       0.54    $^{+    0.15    }_{-    0.13    }$ \\
SFR$_{  radio}$ $z\sim0.7$ &    -0.17   $^{+    0.31    }_{-    0.33    }$ &       0.68    $^{+    0.19    }_{-    0.18    }$ \\

\hline
\end{tabular}\label{tab2}
\end{table}

\subsection{Opacity at UV wavelengths: Tuffs et al. (2004) models}\label{tuffssection}

At UV wavelengths our observations show a clear trend of increasing attenuation with galaxy inclination and a clear evolution of the overall attenuation with redshift. \cite{Tuffs2004} calculate the attenuation of stellar light for a grid of disk opacity and inclination, providing a direct 
comparison to our observations.

\begin{figure*}
\includegraphics[width=0.5\linewidth]{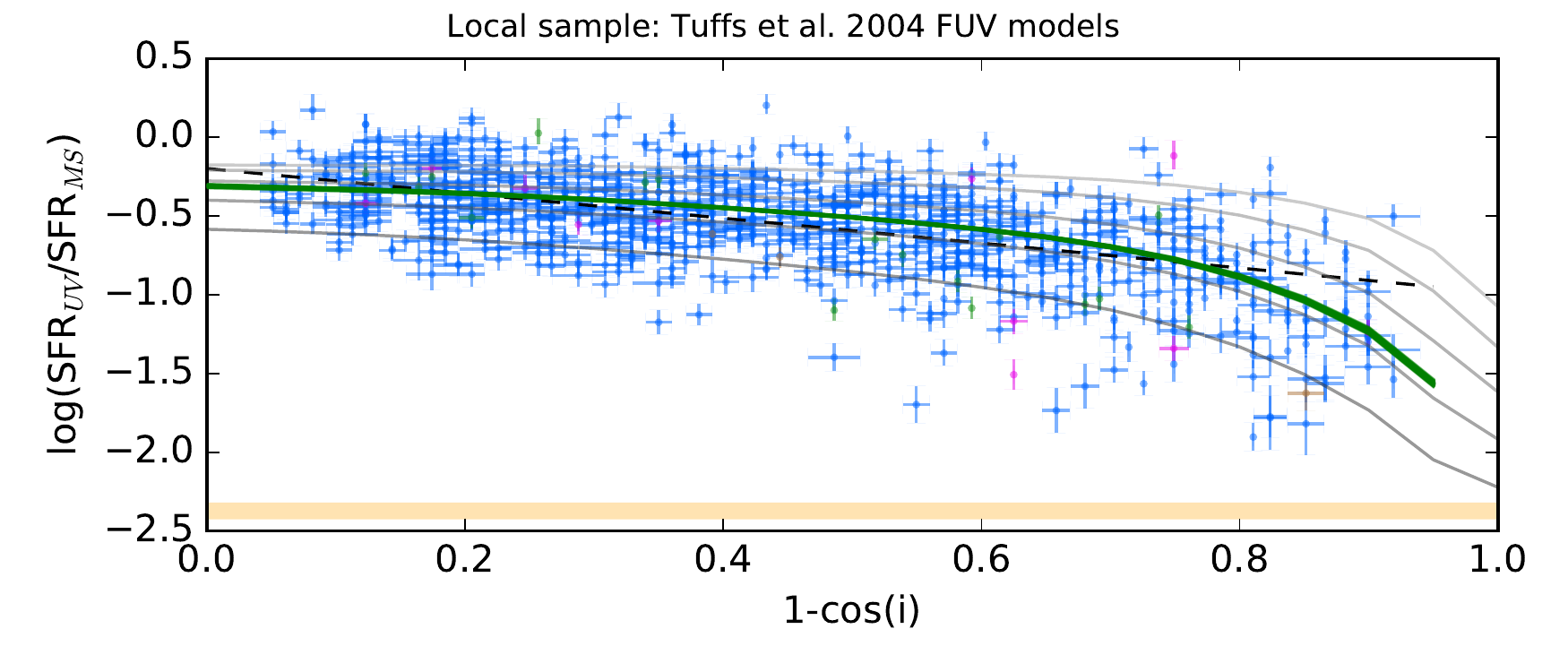}
\includegraphics[width=0.5\linewidth]{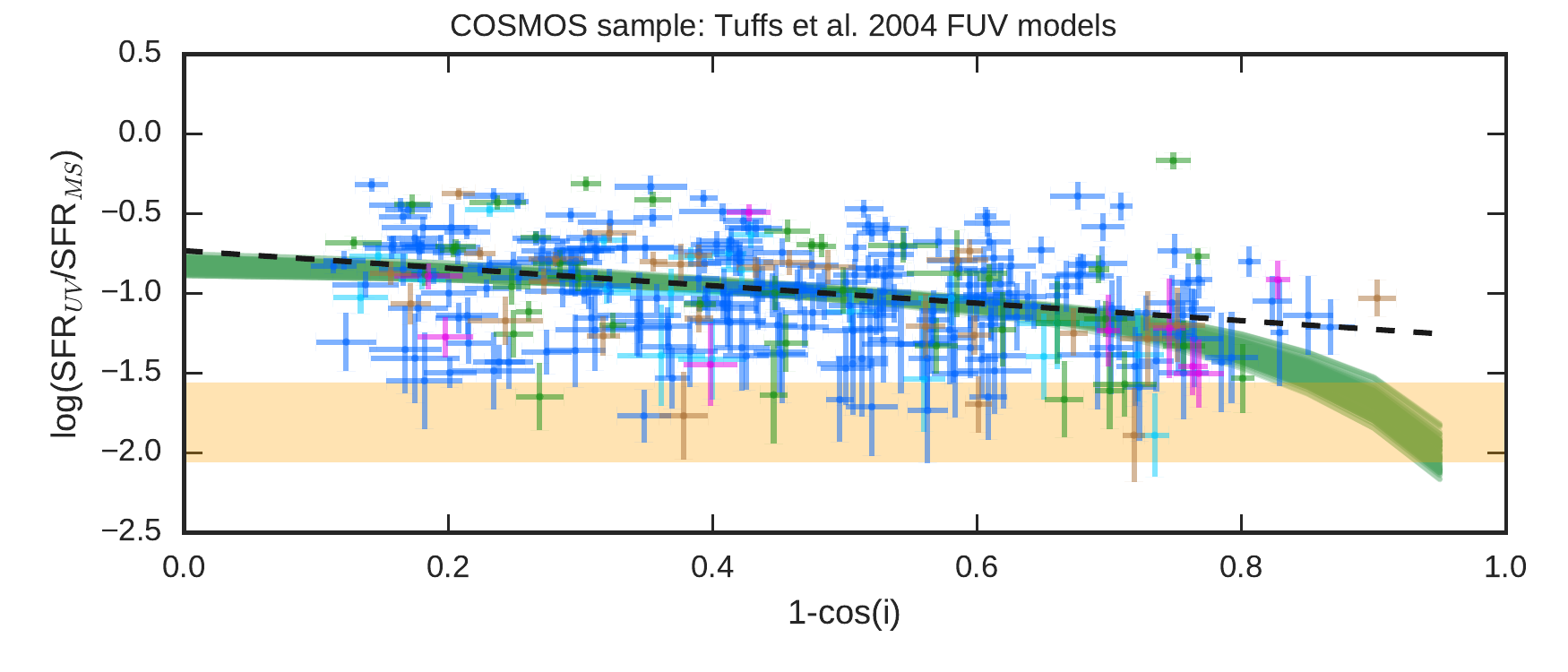}
\caption{SFR$_\text{UV}$ - inclination relation shown in the top panels of Figure 3, but with the attenuation curves of 
Tuffs et al. 2004 showing the expected change in flux, and hence SFR$_{\text{UV}}$, 
with inclination. Green trends show the relations that best fit our data, with 
 $\tau_B^f$ = 3.95$^{+0.16}_{-0.15}$ and 
$F=0.09^{+0.02}_{-0.02}$ for the local sample 
(left) and $\tau_B^f$ = 3.5$^{+1.1}_{-1.8}$,  $F=0.55^{+0.06}_{-0.04}$ for the 
COSMOS sample (right). In the left panel, models shown have face-on B-band opacities of 
$\tau_B^f$= 0.5, 1.0, 2.0, 4.0, and 8.0 (from light to dark grey) and $F=0.22$.}\label{tuffs}
\end{figure*}

\cite{Tuffs2004} calculated the attenuation of stellar light from local spiral galaxies using geometries that are able to reproduce observed galaxy SEDs from the UV to the submillimetre. The model predictions of \cite{Tuffs2004}, which are based on 
the dust model of \cite{Popescu2000}, are shown in Figure \ref{tuffs}. 
The model includes an exponential diffuse dust disk that follows a disk of old 
stars, a second diffuse dust disk that is designed to emulate the dust in spiral 
arms by having the same spatial distribution of the young stellar population 
(the thin disk), a distribution of clumpy, strongly heated grains, correlated 
with star-forming regions, and a dustless de Vaucouleurs bulge. The so-called 
clumpy dust component was required to reproduce the FIR colours and is associated with the opaque parent molecular clouds of the massive stars \citep{Tuffs2004}. 
The diffuse dust disk associated with the spiral arms was needed to account for the sub-millimetre emission.

\cite{Tuffs2004} have calculated attenuation curves (the change of magnitude $\Delta m$ due to dust attenuation) of the thin disk, thick disk, and bulge component ($\Delta m$ vs. 
1-$\cos(i)$) and they have been represented with a fifth-order polynomial at various $\tau_B^f$ and wavelengths, 
where $\tau_B^f$ is the opacity through the centre of a face-on galaxy in the B-band.  Since 
$\tau_B^f$ is the opacity through the centre of the galaxy where most of the dust is concentrated, $\tau_B^f \leq $1.0 corresponds to an optically thin galaxy over most of its area. The attenuation calculations in the UV range were only performed on the thin disk because the thick disk is assumed not to emit in the UV. 

We linearly interpolated the relation for UV attenuation at 1350 and 1650\AA, 
tabulated in \cite{Tuffs2004} for B-band opacities of $\tau_B^f$= 0.5, 1.0, 2.0, 4.0, and 8.0 to 
obtain the expected UV attenuation at 1542\AA, the effective wavelength of the 
GALEX FUV filter. 
We show how the expected UV attenuation affects the SFR$_{\text{UV}}$ using the 
tabulated relation at B-band opacities of $\tau_B^f$= 
0.5, 1.0, 2.0, 4.0, and 8.0 in Figure \ref{tuffs}. 

The attenuation by the clumpy dust component is independent of the inclination of the galaxy, unlike the diffuse dust component. The clumpy dust component is associated with opaque parent clouds of massive stars; no UV light will escape the cloud independent of the viewing angle. The wavelength dependence of the attenuation is also not determined by the optical properties of the grains because the clouds are opaque at each wavelength. Instead, wavelength dependence arises because stars migrate away from their birth cloud over time, and therefore lower-mass, redder stars are more likely to have left their birth clouds and thus their starlight will be less attenuated \citep{Tuffs2004}. 
The \cite{Tuffs2004} models analytically treat the attenuation due to the clumpy component separately, parametrized by the fraction of attenuation by clumpy component, $F$.  

Assuming that only the thin disk emits in the UV range and the disk and bulge 
only in the optical/NIR range as suggested by \cite{Tuffs2004}, we can vary the 
fraction of the emitted FUV flux density, $F_{\text{FUV}}$ , that is locally absorbed in 
the star-forming regions by the clumpy dust component. The total attenuation 
would then be 
\begin{equation}
\Delta m_{\text{UV}} = \Delta m_{\text{UV}}^{\text{tdisk}} -2.5\log\left(1-F_{\text{FUV}} \right)
,\end{equation}
where $\Delta m_{\text{UV}}^{\text{tdisk}}$ is the thin disk attenuation given in 
\cite{Tuffs2004} and $F_{\text{FUV}} = f_{\text{FUV}}\times F$ , where $F$ is the clumpiness fraction 
and $f_{\text{FUV}} = 1.36$ at 1542\AA~ (the wavelength dependence of the probability that a UV photon will be completely absorbed locally).  In this way $F$ will change the overall normalisation of the SFR$_{\text{UV}}$ inclination relation. 

We use the Goodman-Weare Monte Carlo Markov Chain (MCMC) sampling method implemented in \emph{python} package \emph{emcee.py} \citep{ForemanMackey2013} to find posterior distributions of  $\tau_B^f$ and $F$ for our local UV 
galaxy sample. We use an uninformative flat prior between $\tau_B^f$ = [0,10] and $F$= [0,0.61]. The maximum F = 0.61 constraint corresponds to the case where there is a complete lack of cloud fragmentation due to feedback and the probability that photons will escape into the diffuse ISM is determined only by the migration
of stars away from their birth-clouds \citep{Popescu2011}. We keep 800 steps (after a 200 step burn-in) from 50 MCMC walkers. The posterior median values of parameters with 68\% credible intervals for the \cite{Tuffs2004} models at $z\sim0$ are $\tau_B^f$ = 3.95$^{+0.16}_{-0.15}$ and 
$F=0.09^{+0.02}_{-0.02}$. In Figure \ref{tuffs}, we show in green 100 samples drawn randomly from the chain.

The attenuation curves for models with $\tau_B^f$ ranging from 0.5 (light 
grey) to 8 (dark grey) and $F=0.22$ are also shown with our data in Figure 
\ref{tuffs}.
\cite{Tuffs2004} suggested using $F=0.22$ because it was the median value obtained 
from fitting five nearby edge-on spirals, and that value was also obtained for the prototype galaxy NGC 
891.  On the other hand, \cite{Popescu2011} find a clumpiness fraction of $F=0.41$ for typical spiral galaxies in the local universe. Using the FUV, which is most sensitive to the clumpiness factor, we find lower values for $F$ in local galaxies than \cite{Tuffs2004} and \cite{Popescu2011}. However, we are insensitive to dust clouds that are completely obscured in the FUV but still radiate in the infrared. This is likely the reason why the clumpiness factors we measure for the FUV are lower than the SED results that include infrared emission (by a factor of $\sim$2).

\cite{Driver2007} found $\tau_B^f\sim 3.8 \pm 0.7$,
derived from integrated galaxy properties of $\sim$10,000 galaxies with 
bulge-disk decompositions, consistent with our result at $z\sim 0$. At this 
face-on central optical depth, less than half the bolometric luminosity is 
actually absorbed by dust \citep{Tuffs2004}.

For the $z\sim0.7$ sample, we noticed that the fits were dominated by the high-S/N data points with high SFRs, giving preferentially shallower slopes. The S/N effects are more important at $z\sim0.7$ because the detection threshold is close to the data (as opposed to $z\sim 0$, where the detection threshold is well below our data). 
Ignoring the errors on the SFR by setting them all to a constant 0.3 dex, and considering only galaxies with 
$1-\cos(i)<$0.75, where we are more complete in sSFR, we obtain $\tau_B^f$ = 
3.5$^{+1.1}_{-1.8}$ and $F=0.55^{+0.06}_{-0.04}$. Including all data points gives a smaller $\tau_B^f = 2.0^{+0.5}_{-0.3}$ and a correspondingly larger $F = 0.60^{+0.01}_{-0.02}$.

The spread in attenuations in the COSMOS sample could also be partially explained by the fact that galaxies have different clumpy dust fractions. 
\cite{Misiriotis2001} found that $F$ varies by 30\% between galaxies. 

In this analysis, $\tau_B^f$ decreases with redshift while $F$ increases, meaning the overall opacity increases as it is dominated by $F$ at UV wavelengths. The decrease in $\tau_B^f$ is in apparent contradiction with the \cite{Sargent2010} result that reported more opaque galaxies at $z=0.7$ in the B band. However, we note that the errors on $\tau_B^f$ are so large that the decrease reported in this analysis is not statistically significant. Furthermore, we find that normalising the SFR$_{\text{UV}}$ by a different MS produces an increase in $\tau_B^f$ with redshift, while
changing the normalisation did not change the fact that $F$ increases between $z\sim 0$ and $z\sim0.7$, a key result which we discuss in Section 4.

\subsection{UV opacity and stellar mass surface density}

\begin{figure}
\includegraphics[width=\linewidth]{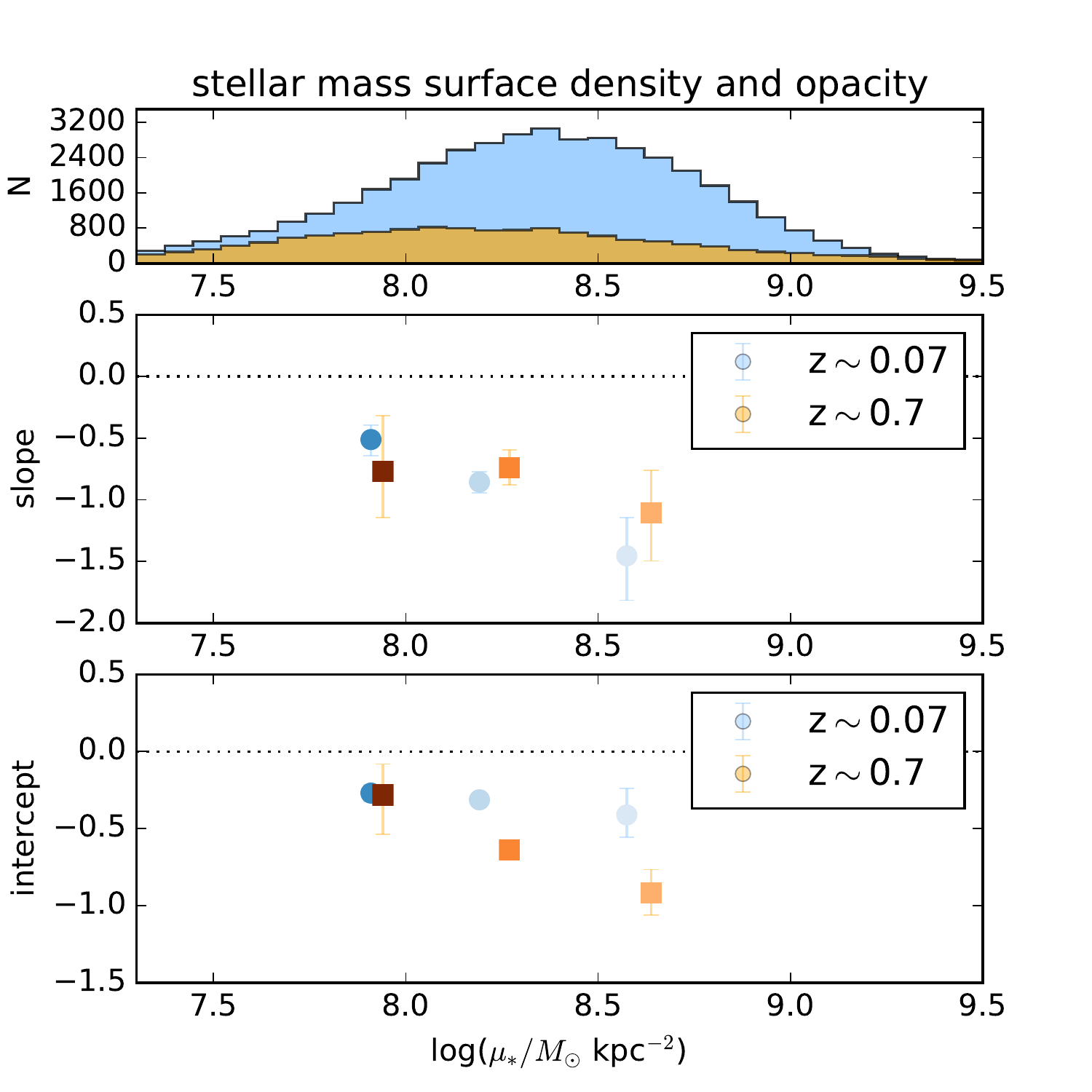}
\caption{Top panel: Distribution of stellar mass surface densities, $\mu_*$, for 
our two redshift ranges. Middle panel: The relation between opacity, represented 
by the slope of the SFR$_{\text{FUV}}$ -inclination relation, and stellar mass surface 
density. Bottom panel: The relation between overall UV attenuation, represented 
by the intercept of the SFR$_{\text{FUV}}$-inclination relation, and $\mu_*$.
Data are colour-coded by the median size, r$_{1/2}$, of each bin in the bottom panel, with colours ranging from 4 kpc (light) to 7 kpc (dark).}\label{mu}
\end{figure}

\cite{Grootes2013} found that the opacity of spiral galaxies in the local 
Universe ($z\leq0.13$) depends on the stellar mass surface density $\mu_*$. 
\cite{DeVis} also report a mass surface density dependence of the B-band 
opacity and suggest that the increased stellar mass potential associated with higher stellar mass surface density creates instabilities in the cold ISM, which lead to the formation of a thin dust disk that increases the attenuation \citep{Dalcanton2004}. 
\cite{Williams2010} show that the stellar mass surface density increases with redshift out to $z\sim2$. We re-compute our SFR$_{\text{FUV}}$-inclination relations in bins of $\mu_*$ to investigate this dependence.
We compute $\mu_*$ for each galaxy using the stellar mass and the physical 
half-light radius from the rest-frame g band GIM2D fits, 
\begin{equation}
\mu_* = \frac{M_*}{2r_{1/2}^2}
.\end{equation}
Figure \ref{mu} shows the distribution of $\log(\mu_*)$ for our samples at $z\sim 0$ and $z\sim0.7$.
We note that $\mu_*$ was calculated with the original stellar mass and half-light 
radius values, without any inclination corrections. Before slicing the samples into bins of $\mu_*$, we corrected $\log(\mu_*)$ for inclination dependence as done in Section 2.3.

Figure \ref{mu} shows that in the local Universe, the slope of the SFR$_{\text{FUV}}$- 
the inclination is steeper for galaxies with higher $\mu_*$ when considering our sample of large and massive galaxies. 
This supports the trend reported in \cite{Grootes2013} that opacity is proportional to $\mu_*$. At $z\sim0.7$, no trend between slope and $\mu_*$ is discernible within the scatter (some bins have scatters smaller than the symbol size).
On the other hand, the intercept of the SFR$_{\text{FUV}}$- inclination relation (the attenuation of face-on disks), clearly decreases (increases) with $\mu_*$ for both samples. 

\cite{Grootes2013} constrained $\tau_B^{f}$ using the FIR SED to measure the dust mass while keeping the clumpiness factor fixed at $F=0.41$. We might 
expect that the IR-derived $\tau_B^f$ of \cite{Grootes2013} should be more closely 
related to our values for the intercept of the  SFR$_{\text{FUV}}$-inclination 
relation. 
For the total attenuation corrections, both the slope and the intercept are important.

We also find that the slope of the SFR$_{\text{UV}}$ -- inclination relation becomes steeper at higher stellar masses at both 
$z\sim 0$ and $z\sim0.7$ (not shown). This indicates that the most massive galaxies are more opaque, and their SFRs are most affected by galaxy inclination.

\section{Evolution of disk opacity}\label{discussion}

\begin{figure*}
\centering
\includegraphics[width=0.9\linewidth]{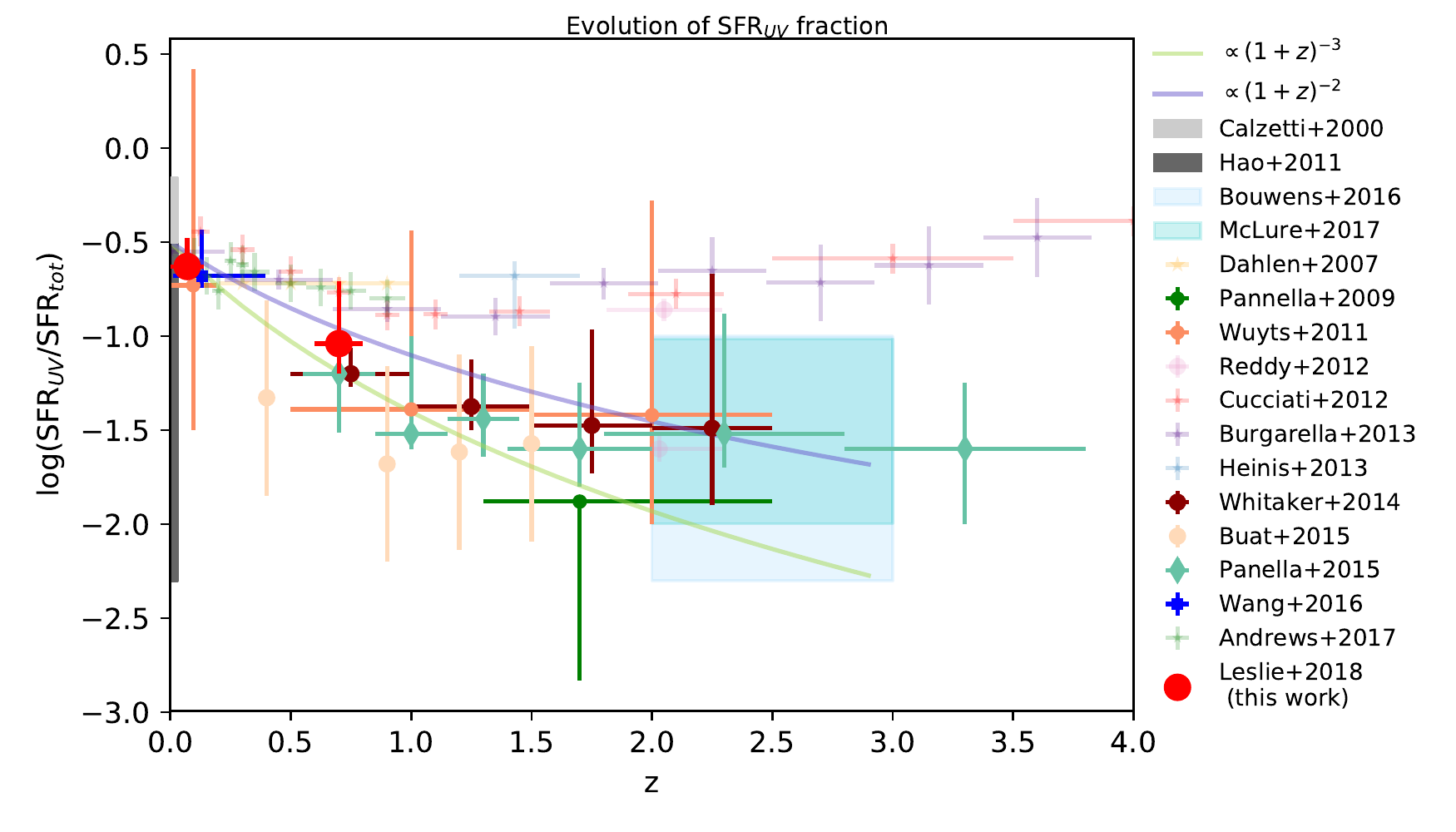}
\caption{Evolution of SFR$_{\text{UV}}$/SFR$_{\text{tot}}$ for statistical samples of galaxies at $z<4$. SFR$_{\text{tot}}$ is the sum of the attenuated (SFR$_\text{IR}$) and non-attenuated (SFR$_{\text{UV}}$) emission. Studies that included mass-dependent measurements are \cite{Wuyts2011b}, \cite{Pannella2009}, \cite{Pannella2015}, \cite{Whitaker2014}, \cite{McLure2017}, and \cite{Bouwens2016}. For these studies, the error-bars show the range of SFR$_{\text{UV}}$/SFR$_{\text{tot}}$ for galaxies with stellar masses
$10.2<\log(M_*/\mathrm{M_\odot})<11.4$, to match the sample of massive galaxies used in this work. \cite{Bouwens2016} and \cite{McLure2017} combined literature results and are shown as shaded areas. Studies that include less-massive galaxies (shown with transparent symbols) have, on average, less attenuated UV emission, so the SFR$_{\text{UV}}$/SFR$_{\text{tot}}$ ratio is higher. 
The red dots show the value of SFR$_{\text{UV}}$/SFR$_\text{MS}$ we obtained for a disk galaxy inclined at $1-\cos(i)=0.5$ ($i = 60\deg$). More information about the studies included in this Figure is provided in Appendix C.
To convert the y-axis to $A_{\text{FUV}}$ we multiply by -2.5. Two solid lines show evolution of SFR$_{\text{UV}}$/SFR$_{\text{tot}}$ between $(1+z)^{-2}$ (purple) and $(1+z)^{-3}$ (lime green) normalised to -0.5 at $z=0$. The evolution of FUV attenuation for massive galaxies evolves as $(1+z)^{2 - 3}$ up to $z\sim2$.
}\label{evolution}
\end{figure*}

Ultra-violet observations are sensitive to dust attenuation, showing trends that other 
wavelengths do not, allowing us to probe disk opacity. We study the UV-derived SFR of disk-dominated galaxies at $z\sim 0$ and $z\sim0.7$ and find a strong inclination dependence that appears to be redshift independent. We infer that the overall FUV attenuation has increased between $z\sim 0$ and $z\sim0.7$ from the collective decreased fraction of SFR$_{\text{UV}}$/SFR$_\text{MS}$ ratios.
To try to understand how the combination of (a) a largely unchanging slope of the UV attenuation-inclination relation, and (b) an evolving overall attenuation might arise we employ the \cite{Tuffs2004} models. 
We find that an increased dust clumpiness can explain most of the increase in attenuation while keeping the slope consistent. 

In the models of \cite{Tuffs2004}, the FUV emission and the clumpy dust component surrounding the HII regions are assumed to only be associated with the thin disk. The fraction of dust mass in the clumpy component is parametrized by $F$ and we find best fitting values 
$F=0.09^{+0.02}_{-0.02}$ and  $F=0.55^{+0.06}_{-0.04}$ for the $z\sim0$ and $z\sim0.7$ 
samples, respectively. We find that the fraction of SFR that is 
attenuated increases by a factor of $\sim$3.4 for face-on galaxies, which corresponds to the increase 
in the viewing-angle-independent clumpy dust fraction $F$ by a factor of $\sim$6
from $z=0.07$ to $z=0.7$. We also find that the average face-on B-band opacity, 
$\tau_f^B$, does not evolve strongly with redshift when traced by the FUV 
emission ($\Delta \tau^f_{B}= -0.45^{+1.1}_{-1.8}$; consistent with zero). 
At $z\sim0.7$, galaxies at fixed stellar mass had more 
mass in gas, and therefore higher SFRs, by factors of 
$\sim$3-4. Stars are born from the dusty molecular clouds and the young stars then heat the dust that is encompassing them; the warm clumpy component thus somewhat traces the recent star formation. An increased clumpy-dust fraction should change the shape of the infrared SED with redshift. Indeed, many studies find that the peaks of the IR SEDs move to shorter rest frame wavelengths with redshift (implying a warmer component in the FIR emission, e.g. \citealt{Magnelli2014, Bethermin2015}).

The star and dust geometry are crucial to understanding how galaxy properties such as observed size $r_{1/2}$ or UV emission are affected by dust attenuation. We note that the geometry of the \cite{Tuffs2004} model may not be valid at $z>0$. We caution that measuring the slope or normalisation of the SFR$_{\text{FUV}}$ -- inclination equation alone does not give a full picture of the dust. For example, a lack of slope could be because a large amount of dust at large scale heights heavily absorbs light even when the galaxy is face-on. Therefore, the slope and the intercept should be considered together for a better understanding of the dust content.
Gas disks are observed to become puffier at higher redshift, having larger scale heights perhaps due to a larger gas fraction and turbulence (e.g. \citealt{Kassin2012, ForsterSchreiber2009, Wisnioski2015}). If the stellar disk height also increases with redshift, then the minimum axis ratio used in calculating inclination (Eq. \ref{eqinc}) must be adjusted accordingly. To date, no studies have conclusively measured the scale height of stellar disks as a function of redshift. However, 
we find no significant difference in the minimum observed axis ratio of our two samples 
(see Fig. 1), suggesting that the stellar disk height has not increased significantly by $z=0.7$. The \cite{Tuffs2004} model assumption of the thick disk not emitting at UV wavelengths might still be a valid assumption at $z\sim0.7$. 
A comparison to simulations that include different dust components, such as those presented recently by \cite{Nelson2017}, or other hydrodynamical and radiative transfer simulations such as \cite{Jonsson2010} or \cite{Trayford2017}, should allow us to test our assumptions about the relative stellar and dust geometries at $z\sim0.7$ and verify whether an increase of the clumpy dust-component is responsible for the increase of the FUV attenuation of massive galaxies with redshift.

In Figure \ref{evolution}, we compare the FUV attenuation as represented by SFR$_{\text{UV}}$/SFR$_{\text{tot}}$ for our two redshift bins with other studies from the literature. 
We find a large variation in SFR$_{\text{UV}}$/SFR$_{\text{tot}}$ when considering galaxies drawn from different studies that have different sample selections.  
Our findings cannot be extrapolated to the general galaxy population. Studies that include less massive galaxies (represented by symbols with higher transparency) tend to show larger SFR$_{\text{UV}}$/SFR$_{\text{tot}}$ ratios, that is, less UV attenuation than samples restricted to more massive galaxies ($\log(M_*/\mathrm{M_\odot})>10.2$). Studies at redshifts $z>1.5$, such as \cite{Pannella2015} and \cite{McLure2017}, find that stellar mass provides a better prediction of UV attenuation than $\beta$, the UV slope itself. \cite{Wuyts2011b} compared their observed SFR$_{\text{UV}}$/SFR$_{\text{IR}}$ with what would be expected from the SFR surface density, assuming a simple model with a 
homogeneously mixed star-gas geometry. They found that more NUV (2800 \AA) 
emission escaped than expected from the simple model at $z>1$, which they suggest could be due to a more patchy geometry in high-redshift galaxies. 
For more information about the data included in Figure \ref{evolution}, please refer to Appendix \ref{studies}.

\begin{figure}
\includegraphics[width = \linewidth]{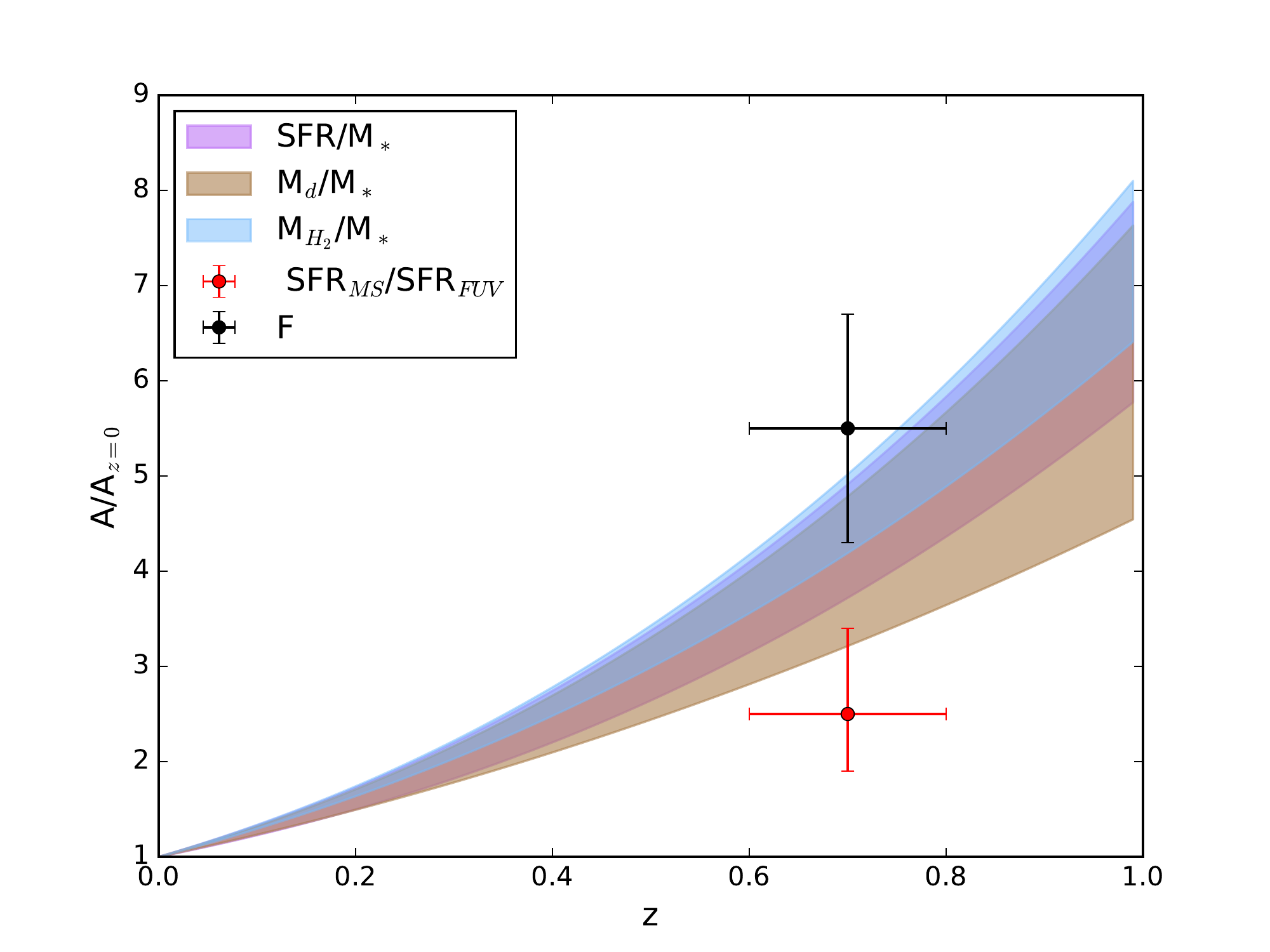}
\caption{Evolution of massive galaxy properties from their local values out to $z\sim$1. The evolution of the dust-mass fraction (M$_d$/M$_*$) was determined from a power-law fit to data from \cite{Santini2014}, \cite{Bethermin2015} (for $z \leq$1), and \cite{RemyRuyer2014} for $z=0$. The upper bound to the M$_\text{d}$/M$_*$ evolution is given by assuming $M_{\text{d}} \sim 0.5 \times Z \times M_\text{g}$, and the evolution of gas and metallicity over this redshift and stellar mass range. The evolution of specific SFR is shaded purple between the relationship in \cite{Sargent2014} and $(1+z)^{3}$. The evolution of the molecular gas fraction (blue) is from \cite{Tacconi2016}. 
We find that the fraction ($F$) of dust in clumps surrounding nascent star-forming regions has increased by a factor of $\sim$5.5, and that the overall attenuation of UV (SFR$_{\text{MS}}$/SFR$_{\text{UV}}$) has increased by a factor of $\sim$2.5 (ranging from 3.4 to 1.9 for face-on to edge-on galaxies, respectively).}\label{evol2}
\end{figure}

We find that the overall FUV attenuation increases with redshift for large, massive star-forming galaxies. 
Interstellar dust is produced in stars which in turn form from the gas; the fact that the dust- and gas-mass fractions vary similarly with redshift is a good consistency check, displayed in Figure \ref{evol2}.
However, the amount of total dust depends on the balance between dust creation and destruction. Keeping track of the interstellar dust that is produced in the ejecta of asymptotic giant branch (AGB) stars and supernovae, the growth of dust grains in dense regions of the ISM, and the destruction of dust in supernovae shocks and collisions is, nevertheless, a difficult task (e.g. \citealt{Aoyama2017, Popping2017}). If we assume, simply, that the increased dust mass with redshift all goes into the clumpy dust component, and that the clumpy component is related to the molecular gas (in terms of amount and distribution) then the amount of overall attenuation could follow the evolution in molecular gas mass. Figure \ref{evol2} shows that the FUV attenuation does not evolve as strongly as the gas mass fraction (as found recently by \cite{Whitaker2017} out to $z\sim2.5$); however, the fraction of dust in the clumpy component ($F$) does evolve strongly with redshift, similar to the gas-mass fraction.

\cite{daCunha2010}, \cite{Smith2012}, 
\cite{Sandstrom2013}, and \cite{Rowlands2014} found that, in the local Universe, M$_\text{d}$/M$_*$ is 
proportional to sSFR. 
Studies such as \cite{Sargent2012} and \cite{Tasca2015}, that look at how sSFR 
changes with redshift, find that the sSFR varies as sSFR$\propto(1+z)^{2.8}$ 
 out to redshifts $z<2$. This would give a sSFR increase by a factor of 
$\sim3.6$ between $z=0.07$ and $z=0.7$. We note, however, that this trend also is stellar
mass dependent (e.g. \citealt{Schreiber2015}), and that lower-mass galaxies have a 
shallower evolution in sSFR than the most massive galaxies. 

Recent studies of molecular gas at high redshifts indicate that the cosmic density of molecular gas, similar to the cosmic density of star formation, peaks between redshifts 2 and 3 \citep{Decarli2016}. \cite{Genzel2015}, \cite{Sargent2014}, and \cite{Tacconi2016} used compilations of studies that 
measure gas mass via CO line luminosities and found that M$_{\text{g}}$/M$_*$ varies like 
$(1+z)^{2.7}$, which corresponds to an increase by a factor of $\sim$3.5 between 
$z=0.07$ and $z=0.7$.
On the other hand, \cite{Bauermeister2013} expect M$_{\text{g}}$/M$_*$ $\propto (1+z)^{3.2}$ which would imply an increase in the gas mass fraction by a factor of $\sim$4.4. 
In any case, the gas mass is expected to increase by a factor of $\sim$4 between our two redshift slices. In Figure \ref{evolution}, we show with two solid lines the relations SFR$_{\text{UV}}$/SFR$_{\text{tot}} \propto (1+z)^{-2}$ and $\propto(1+z)^{-3}$.

The parameter that most strongly controls 
the dust-to-gas mass is the metallicity (e.g. \citealt{Leroy2011, 
Accurso2017}). According to the parametrization of \cite{Zahid2014}, a galaxy with 
log(M$_*$/M$_\odot)=10.2$ would have a metallicity of 0.069 dex less at $z=0.7$ 
than at $z= 0.07$. A galaxy at M$_* \sim$ 10$^{10.5}$M$_\odot$, on the other hand, will have a change in metallicity of 0.035 dex between the same two redshifts. 
For our sample, the dust to gas ratio should not change significantly between $z\sim0$ and $z\sim0.7$, however, we take this change into account in Figure \ref{evol2} when calculating the expected M$_\text{d}$/M$_*$ evolution based on the dust-to-gas ratio.
%


To summarise, out to $z<2$, sSFRs increase as $(1+z)^{2.8}$ and M$_\mathrm{g}$/M$_*$ and 
M$_\text{d}$/M$_*$ increase at a similar rate for galaxies more massive than $10^{10}$ M$_\odot$
\citep{Santini2014}. Since $z=0.7$, SFRs and gas and dust masses have decreased by 
factors of $\sim$3.5.
We see a larger systematic FUV attenuation in the $z\sim0.7$ sample compared to the 
$z\sim 0$ sample. The SFR$_{\text{FUV}}$ of a face-on galaxy lies $\sim$0.21 dex below 
the MS at $z=0.07$, and $\sim$0.74 dex below the MS at $z=0.7$.  Assuming the main-sequence SFR to be the unobscured SFR, or 
SFR$_{\text{UV}}$+SFR$_{\text{IR}}$,
the FUV  attenuation of face-on galaxies has thus increased by a factor of $\sim$3.4.

Galaxies are thought to grow in size over time (`inside-out growth'). For example, \cite{Wuyts2011b} shows that galaxies on the main sequence at a fixed stellar mass have a larger size on average at $z\sim 0.1$ than at $z\sim$1. Due 
to this evolution, and the fact that opacity also varies with radius in galaxies 
out to at least $z<2$ \citep{Tacchella2017, Wang2017}, we cannot interpret the slopes of the attenuation-inclination relation reported in this paper in the 
context of an evolutionary picture for particular galaxies. Instead, we emphasise that the 
inclination dependence of the UV attenuation in the largest and most massive 
star-forming disk galaxies at $z\sim0.7$ (which are not necessarily the same 
galaxies that will end up in our sample at $z=0.07$), behaves similarly to that in 
galaxies selected to have the same stellar mass and radius at $z\sim 0$. However, 
the overall FUV attenuation has increased, presumably due to the larger amount 
of dust at higher redshifts.

The importance of inclination-dependent FUV SFRs, even when common attenuation corrections have been applied, will be the subject of Paper II.

\section{Conclusions}
We parametrize the attenuation of starlight as a function of galaxy disk inclination to investigate the global dust properties by fitting a simple linear function to the SFR -- inclination relation for a sample of galaxies at $z\sim 0$ 
(in SDSS) and $z\sim0.7$ (in COSMOS). We compiled two samples of galaxies at 
four different wavelengths: FUV, MIR (12 $\mu$m), FIR (60 $\mu$m) and radio (20 cm) 
with well measured optical morphology for which we calculated monochromatic 
SFRs. Our findings can be summarised as follows. Massive disks ($\log(M_*/\mathrm{M_\odot}) >10.2$) larger than 4 kpc, show strong FUV SFR--inclination trends at both $z\sim 0$ and $z\sim0.7$. Our fit to this relation is sensitive to selection effects, in particular, to flux limits and size cuts. The FUV SFR of face-on galaxies at $z\sim0.7$ is less than that expected  from the main-sequence relation by a factor of 0.74$\pm$0.09 dex, compared to 
0.20$\pm$0.03 dex at $z\sim 0$. This corresponds to an increase of the FUV attenuation of face-on galaxies by a factor $\sim$3.4 over the last 7.5 Gyr. An increased fraction of dust in warm clumpy components surrounding the 
HII regions (by a factor of about 6) could explain this increased overall attenuation while simultaneously allowing the opacities (the slope) to remain constant.
This increase in the warm clump component and UV attenuation between $z\sim0$ and $z\sim0.7$ is consistent with the increased molecular gas content in galaxies at redshift 0.7. 
Overall FUV attenuation increases with stellar mass surface density at both $z\sim 0$ and $z\sim0.7$.
It is likely that no opacity is present in the FIR and radio, however, we were unable to confirm this for our FIR and radio subsamples as our data did not span a sufficient range in luminosity due to survey flux limits. 
MIR and radio SFRs are inclination independent and therefore MIR and radio data can provide a useful tracer of SFRs also for galaxies with a high inclination angle.

The increase in FUV attenuation for massive galaxies follows the amount of evolution in sSFR, gas, and dust mass fractions over the past 7.5 billion years. These phases (gas, dust and star formation) are therefore closely related, probably even spatially related; an assumption which is used for deriving gas masses from dust continuum emission and for energy balancing SED codes that assume the dust emission and star formation come from similar regions.

\begin{acknowledgements}

MTS was supported by a Royal Society Leverhulme Trust Senior Research Fellowship 
(LT150041).
Based on data products from observations made with ESO Telescopes at the La
Silla Paranal Observatory under ESO programme ID 179.A-2005 and on
data products produced by TERAPIX and the Cambridge Astronomy Survey
Unit on behalf of the UltraVISTA consortium.
This research has made use of the SVO Filter Profile Service 
(http://svo2.cab.inta-csic.es/theory/fps/) supported by the Spanish MINECO 
through grant AyA2014-55216.

AllWISE makes use of data from WISE, which is a joint project of the University of California, Los Angeles, and the Jet Propulsion Laboratory/California 
Institute of Technology, and NEOWISE, which is a project of the Jet Propulsion 
Laboratory/California Institute of Technology. WISE and NEOWISE are funded by the National Aeronautics and Space Administration.

This research has made
use of Astropy13, a community-developed core Python package
for Astronomy (Astropy Collaboration et al., 2013).
This research has made use of NumPy(Walt et al.,
2011), SciPy, and MatPlotLib (Hunter, 2007).
This research
has made use of TOPCAT (Taylor, 2005), which was initially
developed under the UK Starlink project, and has since been supported
by PPARC, the VOTech project, the AstroGrid project, the
AIDA project, the STFC, the GAVO project, the European Space
Agency, and the GENIUS project. 

\end{acknowledgements}

%
\bibliographystyle{mnras.bst} 
 \bibliography{inclination} 
%

\appendix 

\section{Choice of normalisation}
Systematic uncertainties in our analysis arise from the normalisation factor adopted to represent the expected `true' SFR of our galaxies. Here we discuss how our results change as a result of different choices for normalisation; we do not account for all possible differences, but rather focus on a few cases.

\subsection{Choice of main sequence}

\begin{figure}
\includegraphics[width =\linewidth]{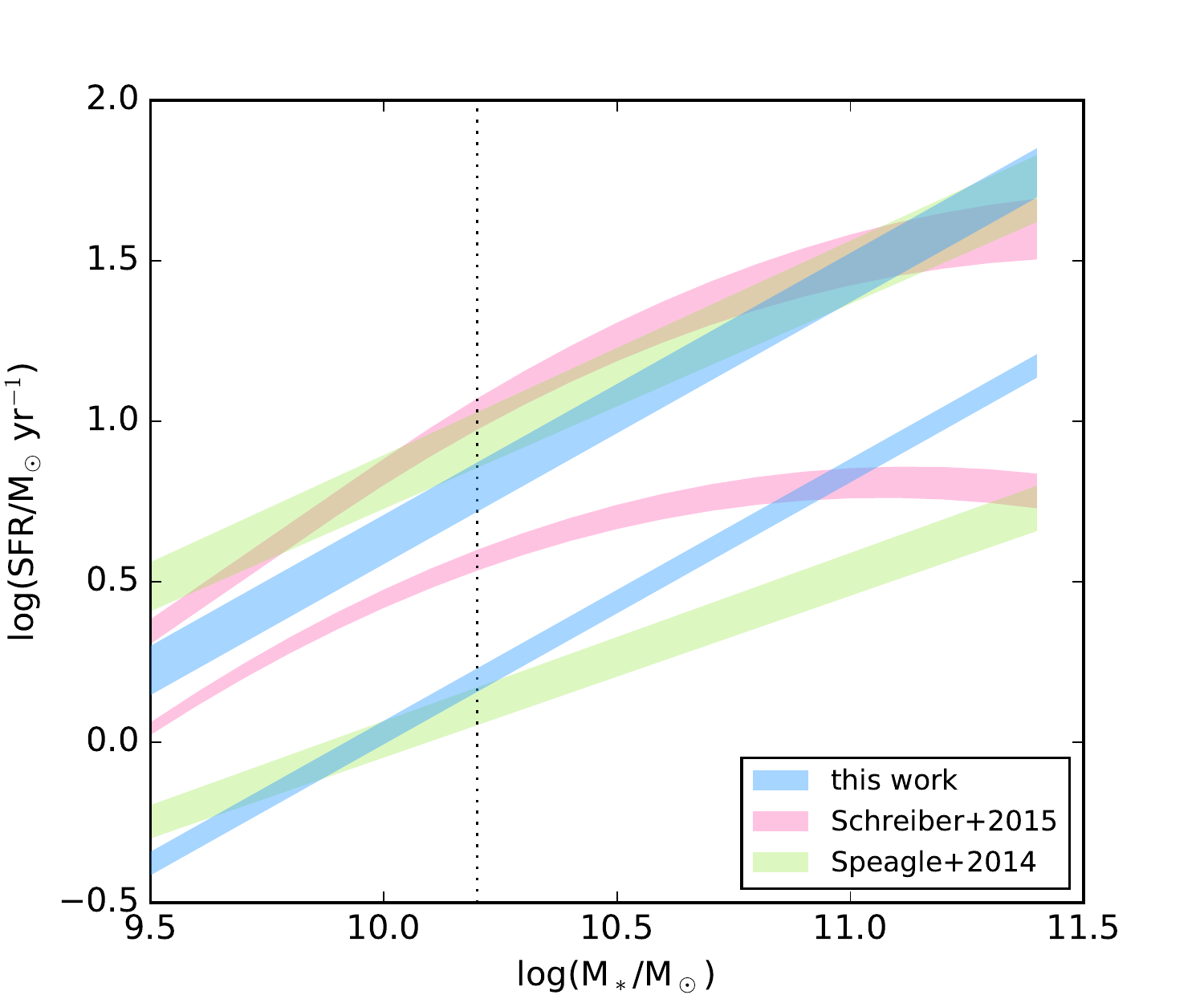}
\caption{Main sequence relations at our two redshift slices. For our analysis we adopt the MS given by Equation 9 (blue), but we also compare to MS relations with a shallower slope, \cite{Speagle2014} (green) and a MS relation with a turn-over at low redshifts and high stellar masses, \cite{Schreiber2015}. The vertical dotted line at $\log(M_*/\mathrm{M_\odot}) = 10.2$ shows the mass cut used in this work. We note that the \cite{Speagle2014} and \cite{Schreiber2015} relations shown for comparison were not calibrated at $z=0$.}\label{MSvisual}
\end{figure}

Many different MS relations exist in the literature. Different methods for selecting ``star-forming'' galaxies tend to produce different results \citep{Karim2011, Speagle2014}. Other differences such as assumed IMF, SFR calibration, and different SED inputs such as star formation histories also make comparisons between studies challenging.  
Studies such as \cite{Schreiber2015}, \cite{Lee2015} and \cite{Whitaker2014}, report a turnover or flattening in the galaxy 
main sequence at high-stellar mass (log($M_*$/M$_\odot$)$>10.3$). The origin of the turnover could be due to samples that include more bulge-dominated 
galaxies, which lie below the main sequence that is defined for pure disk-dominated galaxies (e.g. \citealt{Salmi2012, Lang2015, Whitaker2015}). Figure \ref{MSvisual} is a visualisation of different MS relations from the literature, including the relations used in this work, at $z\sim 0$ (the relations with lower SFRs) and at $z\sim0.7$ (higher SFRs). 
We investigate how our results change when using different main-sequence relations.



\cite{Speagle2014} compiled 25 studies from the literature and found a consensus MS relation after converting all observations to a common set of calibrations. 
We use the best fit 
\begin{equation}
\log(\text{SFR}_\text{MS}) = (0.84-0.026t)\log(M_*) - (6.51-0.11 t)
\label{speagle},\end{equation}
where $t$ is the age of the universe in Gyr. \cite{Speagle2014} removed data with  $t<$2.5 Gyr (and $t>11.5$ Gyr) from their analysis. Their reasons for removing the low-redshift data is because most local studies have been based on the SDSS and most require some aperture corrections. In Figure \ref{MSvisual}, we show Equation \ref{speagle} extrapolated to the redshift of our local sample, $0.04<z<0.1$ is shown in shaded green. 

\cite{Schreiber2015} use a method called ``scatter stacking'' and combine direct UV and FIR light for a mass-complete sample of star-forming galaxies and find a close-to-linear slope of the relation but allow for an observed flattening of the MS that takes places at masses $\log(M_*/\text{M}_\odot)>10.5$ at low redshift. 
\cite{Schreiber2015} found the following equation to represent the locus of the MS:
\begin{multline}
\log(\text{SFR}_\text{MS}) = m - 0.5 +1.5\log(1+z) \\- 0.3\left[\text{max}(0, m-0.36 -2.5 \log(1+z))\right]^2, 
\label{schreiber}\end{multline}
where $m=\log(M_*/10^9\text{M}_\odot)$.
The error bars on Equations \ref{speagle} and \ref{schreiber} were omitted because we did not incorporate them in our SFR-inclination analysis. Figure \ref{MSvisual} shows the \cite{Schreiber2015} MS relation at the redshifts considered for this analysis. The lowest redshift considered in the \cite{Schreiber2015} analysis was $z=0.3$, therefore the low-redshift relation is also an extrapolation of the data. 

For our main analysis, we adopted an updated MS relationship based off an inter-sample `concordance' analysis by \cite{Sargent2014}, which unlike the compiled relation by \cite{Speagle2014}, includes local studies such as \cite{Chang2015} for constraining the MS evolution. 

Table \ref{appendixtable} gives the alternative results for our analysis in Section 3 normalising to these alternative MS equations.
The fitted slope of the SFR$_\text{UV}$/SFR$_\text{MS}$ versus inclination of our local or $z\sim0.7$ samples relation does depend significantly on the MS normalisation selected. On the other hand, the intercept values are different. For the $z\sim0.7$ sample, the intercepts are lower than our adopted MS result for the Speagle and lowest for the Schreiber normalisation. For our $z\sim0$ sample, the intercept is lowest for the Schreiber MS, and highest for the Speagle MS. These results are not surprising considering the normalisation differences seen the MS relations in Figure \ref{MSvisual}. 


Using the \cite{Speagle2014} MS for normalisation of the SFR$_{\text{UV}}$ gives best fitting parameters for the \cite{Tuffs2004} models
$\tau_{B,z=0}^f = 3.19^{+0.1}_{-0.08}$, $F_{z=0} = 0.02 ^{+0.01}_{-0.01}$ for the local galaxy sample, and $\tau_{B,z=0.7}^{f} = 4.6 ^{+1.3}_{-1.5}$, F$_{z=0.7} =0.57^{+0.04}_{-0.03}$ for our COSMOS sample.

On the other hand, using the \cite{Schreiber2015} MS results in $\tau_{B,z=0}^f = 2.89^{+0.1}_{-0.1}$, $F_{z=0} = 0.48 ^{+0.005}_{-0.006}$ and $\tau_{B,z=0.7}^{f} = 5.4 ^{+0.8}_{-1.2}$, F$_{z=0.7} =0.58^{+0.02}_{-0.03}$.

In these cases, the face-on B-band optical depth increases from $z\sim 0$ to $z\sim0.7$ as well as the clumpiness factor (which increases by a factor of $\sim$28 for the \citealt{Speagle2014} MS, and by a factor 1.2 for the \citealt{Schreiber2015} MS). 

\subsection{Attenuation-inclination relations normalised by galaxy SFR$_{\text{MIR}}$}

We investigate the effects that normalising the UV, radio and FIR SFRs by the MIR SFR (12$\mu$m at $z\sim 0$ and 24$\mu$m at $z\sim0.7$), rather than the MS SFR, has on our analysis. 
By using SFR$_{\lambda}$/SFR$_{\text{MIR}}$ for our inclination analysis, we remove potential systematic effects related to the M$_*$ dependence of the star-forming MS (as discussed above). 
SFR$_{\text{MIR}}$ has the practical advantage as the normalising factor (over e.g. SFR$_{\text{FIR}}$) in that most of our SDSS and COSMOS galaxies are detected in the WISE and MIPS data-sets, respectively. 

Attenuation-inclination relations derived using this alternative normalisation are consistent with our results obtained using the MS normalisation, with the exception of a few cases which we will now discuss. 

The slightly positive slope (0.23$\pm$0.21) of the SDSS radio sample is partially due to the fact that the local SFR$_{12\mu\text{m}}$ has a slight negative trend with inclination (-0.12$\pm$0.02), and so normalising by the 12 $\mu$m SFR will induce a slight positive slope. 
The distributions of the FUV slopes now have consistent medians. This could also be in part because the local MIR slope is slightly negative so normalising by the 12 $\mu$m SFR will shift the local UV slope. However, this shift of $\sim$0.12 would not be enough to fully explain the change in local UV slopes (from -0.78$\pm$0.08 to -0.59$\pm$0.1) when using a MS or SFR$_{12\mu\text{m}}$ normalisation. Additionally, the COSMOS SFR$_\text{UV}$ slope changes from -0.53$\pm$0.17 to -0.6$\pm$0.23. 
The local 60 $\mu$m intercept is greater than zero. This corresponds to a difference between the 60 $\mu$m and 12 $\mu$m SFR calibrations that we discuss in Appendix B. 

SFR/SFR$_{\text{MS}}$ will vary from galaxy to galaxy as there is an intrinsic scatter to the MS \citep{Speagle2014,Schreiber2015}. We find that the scatter of the SFR$_{\text{UV}}$/SFR$_{\text{MS}}$-inclination relation is similar to the scatter of the SFR$_{\text{UV}}$/SFR$_{12\mu\text{m}}$ relation (in fact the SFR$_{\text{UV}}$/SFR$_{12\mu\text{m}}$ ratio shows larger scatter). This is not surprising because the amount of UV emission that escapes will depend, amongst other things, on the star-dust geometry, inclination angle, and the dust mass as well as the SFR. Therefore the SFR$_{12\mu\text{m}}$ is not readily correlated to the SFR$_{\text{UV}}$, even for our sample that has been selected to have similar $M_*$, $n$, and $r_{1/2}$.

\begin{table*}
\caption{Best fit parameters of the SFR vs. inclination relation under different SFR normalisations. Column 1 is a repeat of Table 3 (to facilitate comparison), Column 2 and 3 show the best fitting parameters obtained when using the \cite{Speagle2014} and \cite{Schreiber2015} MS relations to normalise the SFR. Column 4 shows the best-fit parameters when the SFR of each galaxy is normalised by its MIR SFR.}\label{appendixtable}
\def\arraystretch{1.5}
\centering
\begin{tabular}{|c|cc|cc|cc||cc|}
\hline
$\lambda$ & \multicolumn{2}{|c|}{This MS} & \multicolumn{2}{|c|}{Speagle MS} & \multicolumn{2}{|c||}{Schreiber MS}& \multicolumn{2}{|c|}{MIR}\\
\hline
 & Slope & Intercept & Slope & Intercept & Slope & Intercept & Slope & Intercept \\
 \hline 
SFR$_{  UV}$    $z\sim 0$ &     -0.78   $^{+    0.08    }_{-    0.09    }$ &       -0.20   $^{+    0.04    }_{-    0.03    }$ &    -0.76   $^{+    0.07    }_{-    0.08    }$ &       -0.11   $^{+    0.03    }_{-    0.03    }$ &    -0.75   $^{+    0.08    }_{-    0.08    }$ &       -0.54   $^{+    0.03    }_{-    0.03    }$ & -0.59$^{+0.10}_{-0.09}$ & -0.27$^{+0.03}_{-0.03}$\\
SFR$_{  MIR}$   $z\sim 0$ &     -0.12   $^{+    0.03    }_{-    0.02    }$ &       0.05    $^{+    0.01    }_{-    0.01    }$ &    -0.10   $^{+    0.02    }_{-    0.02    }$ &       0.14    $^{+    0.01    }_{-    0.01    }$ &    -0.08   $^{+    0.02    }_{-    0.03    }$ &       -0.29   $^{+    0.01    }_{-    0.01    }$ & - & -\\
SFR$_{  FIR}$   $z\sim 0$ &     -0.06   $^{+    0.22    }_{-    0.20    }$ &       0.68    $^{+    0.09    }_{-    0.09    }$ &    -0.02   $^{+    0.19    }_{-    0.18    }$ &       0.81    $^{+    0.08    }_{-    0.08    }$ &    0.04    $^{+    0.19    }_{-    0.17    }$ &       0.40    $^{+    0.08    }_{-    0.08    }$ & 0.05$^{+0.16}_{-0.16}$ & 0.26 $^{+0.08}_{-0.08}$\\
SFR$_{  radio}$ $z\sim 0$ &     -0.10   $^{+    0.20    }_{-    0.22    }$ &       0.52    $^{+    0.11    }_{-    0.09    }$ &    -0.09   $^{+    0.20    }_{-    0.17    }$ &       0.69    $^{+    0.09    }_{-    0.09    }$ &    -0.07   $^{+    0.16    }_{-    0.19    }$ &       0.31    $^{+    0.10    }_{-    0.09    }$ & 0.23$^{+0.21}_{-0.21}$ & -0.07$^{+0.12}_{-0.24}$\\
\hline
SFR$_{  UV}$    $z\sim0.7$ &    -0.54   $^{+    0.18    }_{-    0.17    }$ &       -0.74   $^{+    0.08    }_{-    0.08    }$ &    -0.59   $^{+    0.17    }_{-    0.18    }$ &       -0.85   $^{+    0.09    }_{-    0.07    }$ &    -0.59   $^{+    0.17    }_{-    0.17    }$ &       -0.94   $^{+    0.08    }_{-    0.08    }$ & -0.60$^{+0.23}_{-0.21}$ & -1.00$^{+0.09}_{-0.1}$\\
SFR$_{  MIR}$   $z\sim0.7$ &    -0.03   $^{+    0.10    }_{-    0.09    }$ &       0.28    $^{+    0.05    }_{-    0.05    }$ &    -0.05   $^{+    0.10    }_{-    0.10    }$ &       0.16    $^{+    0.06    }_{-    0.05    }$ &    -0.04   $^{+    0.09    }_{-    0.10    }$ &       0.06    $^{+    0.05    }_{-    0.05    }$ & - & -\\
SFR$_{  FIR}$   $z\sim0.7$ &    -0.02   $^{+    0.23    }_{-    0.25    }$ &       0.54    $^{+    0.15    }_{-    0.13    }$ &    -0.03   $^{+    0.23    }_{-    0.21    }$ &       0.43    $^{+    0.12    }_{-    0.13    }$ &    -0.03   $^{+    0.22    }_{-    0.23    }$ &       0.33    $^{+    0.13    }_{-    0.12    }$ & 0.16 $^{+0.1}_{-0.14}$ & -0.02$^{+0.08}_{-0.11}$\\
SFR$_{  radio}$ $z\sim0.7$ &    -0.17   $^{+    0.31    }_{-    0.33    }$ &       0.68    $^{+    0.19    }_{-    0.18    }$ &    -0.14   $^{+    0.27    }_{-    0.32    }$ &       0.55    $^{+    0.18    }_{-    0.15    }$ &    -0.13   $^{+    0.30    }_{-    0.30    }$ &       0.45    $^{+    0.18    }_{-    0.15    }$ & 0.29 $^{+0.18}_{-0.20}$ & -0.05$^{+0.11}_{-0.10}$\\

\hline
\end{tabular}
\end{table*}

\section{Comparison of SFR tracers}

\begin{figure*}
\includegraphics[width=0.5\linewidth]{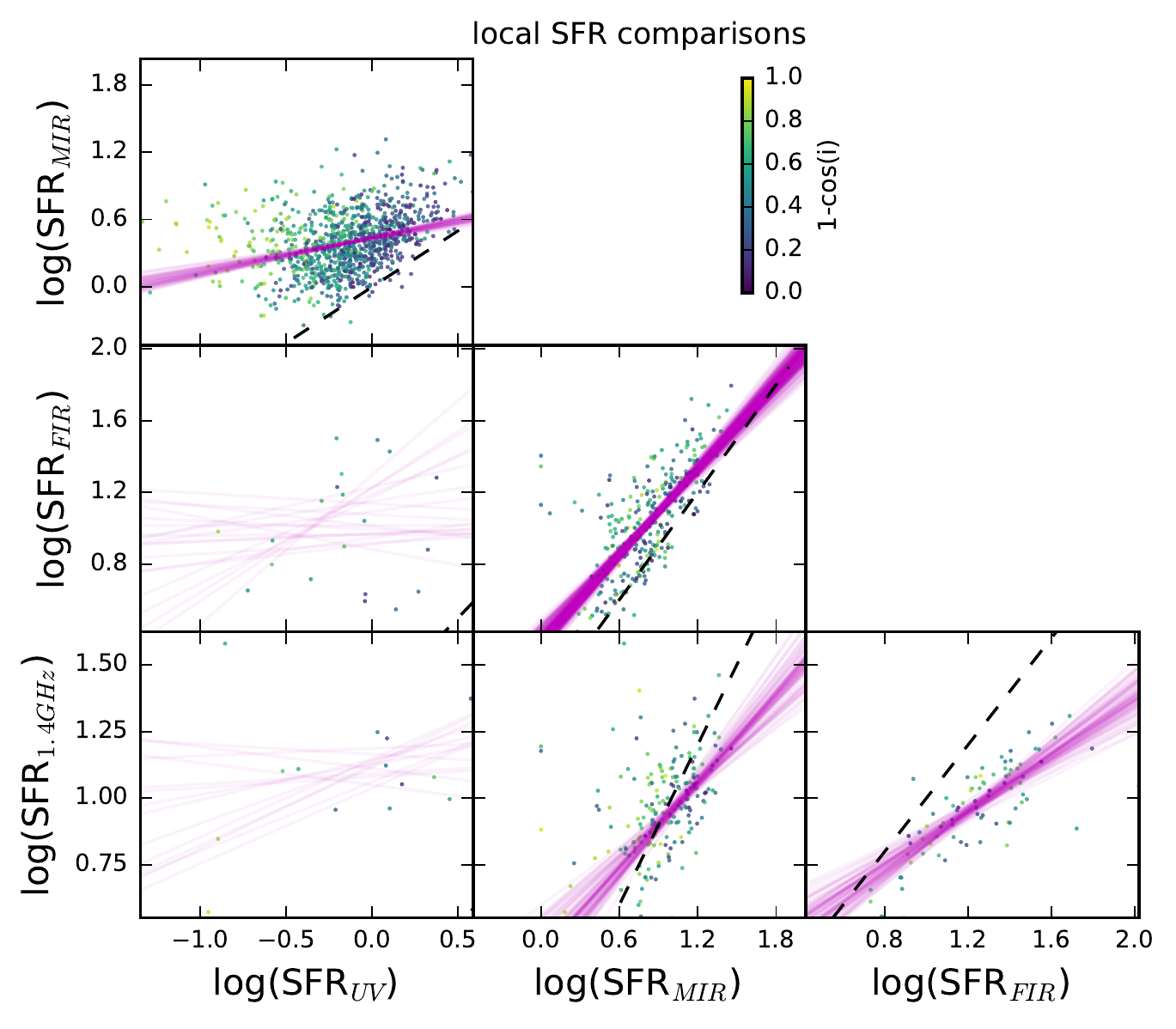}
\includegraphics[width=0.5\linewidth]{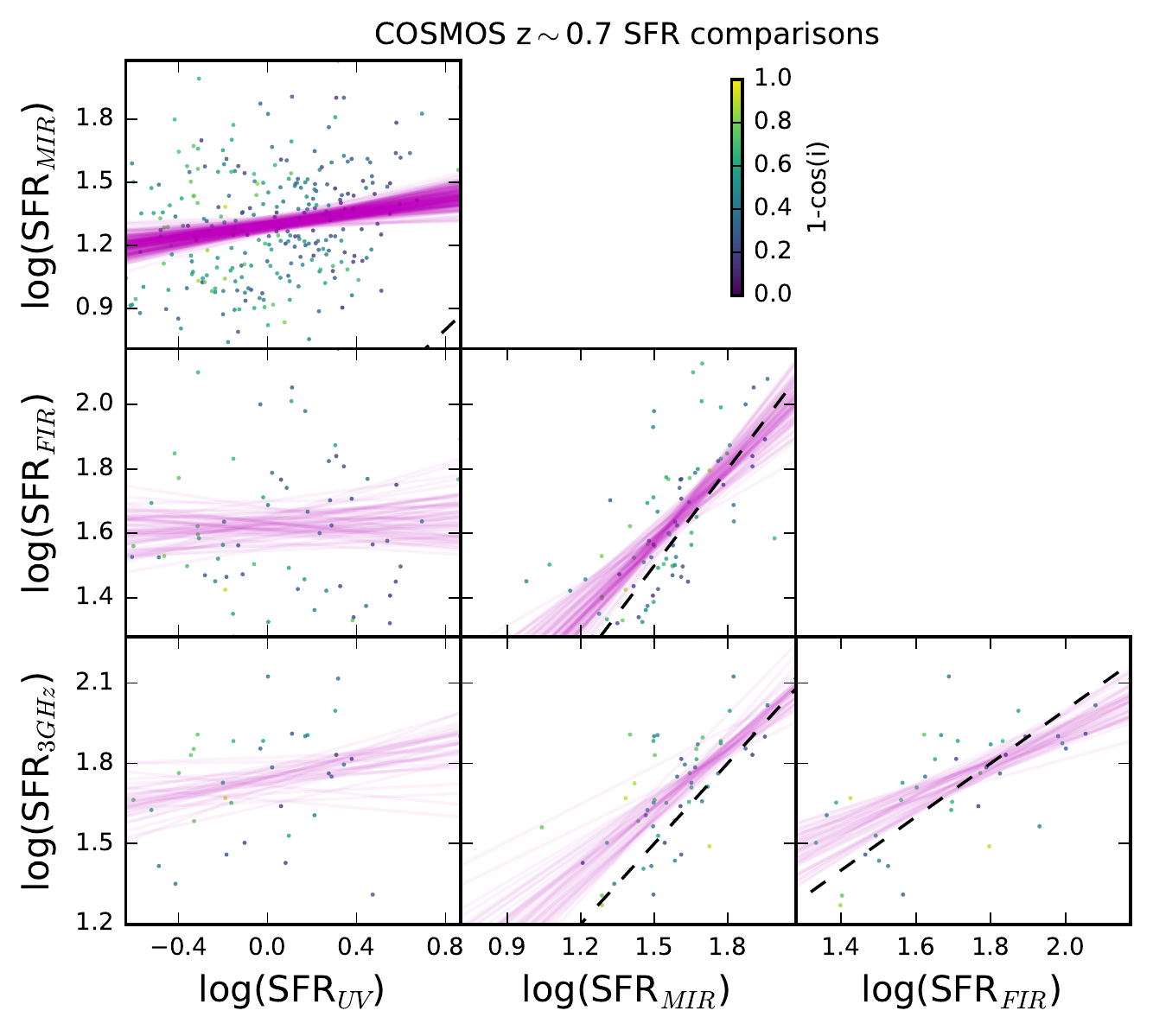}
\caption{Comparison of the SFRs derived using the different monochromatic calibrations used in this paper and described in Section 2.5. The dashed line represents the one-to-one line. Magenta lines show best fit linear relations to resampled data, with the number of fits shown equal to the number of galaxies in each panel. The SFR$_{\text{UV}}$ is not corrected for dust attenuation and so is expected to be smaller than the other indicators. Data are colour-coded by galaxy inclination angle. }\label{sfrcomp}
\end{figure*}

Figure \ref{sfrcomp} shows how the multiwavelength SFRs of galaxies used for our analysis compare in the cases where galaxies are detected in the relevant bands. 
It is important to recognise that these different SFR tracers come from emission produced in different regions (from the HII regions, dust, supernovae remnants) and trace processes on different timescales. FUV, TIR, 24 $\mu$m and 1.4 GHz emission come from stellar populations 0-100 Myr old \citep{Kennicutt2012}, but the values are sensitive to the star formation history. 

We fit a linear function to relate the logarithmic SFRs using the same fitting and resampling techniques used in Section 3. However, in this case, we re-sample the data $N$ times, where $N$ is the number of galaxies in our sample detected at both wavelengths. The best fit lines from each re-sampled data-set are shown as magenta lines in Figure \ref{sfrcomp}. We report the median and error bars (given by the 5th and 95th percentile) in the equations below. The numbers in the brackets show the number of galaxies available for the fit. 

\begin{align*}
\log(\text{SFR}_{\text{UV}}) &= 0.31^{+0.04}_{-0.06}\log(\text{SFR}_{\text{MIR}}) +0.44^{+0.01}_{-0.01} & (1019)\\
\log(\text{SFR}_{\text{UV}}) &= 0.1^{+0.1}_{-0.5}\log(\text{SFR}_{\text{FIR}}) +1.0^{+0.1}_{-0.2} &(20)\\
\log(\text{SFR}_{\text{UV}}) &= 0.2^{+0.4}_{-0.1}\log(\text{SFR}_{1.4\text{GHz}}) +1.09^{+0.05}_{-0.07} & (14)\\
\log(\text{SFR}_{\text{MIR}}) &= 0.79^{+0.08}_{-0.08}\log(\text{SFR}_{\text{FIR}}) +0.38^{+0.07}_{-0.07} & (290)\\
\log(\text{SFR}_{\text{MIR}}) &= 0.56^{+0.15}_{-0.07}\log(\text{SFR}_{1.4\text{GHz}}) +0.4^{+0.1}_{-0.1} & (83) \\
\log(\text{SFR}_{\text{FIR}}) &= 0.53^{+0.07}_{-0.10}\log(\text{SFR}_{1.4\text{GHz}}) +0.3^{+0.1}_{-0.1} & (83)\\
\end{align*}

COSMOS z$\sim$ 0.7 results: 
\begin{align*}
\log(\text{SFR}_{\text{UV}}) &= 0.15_{-0.08}^{+0.09} \log(\text{SFR}_{\text{MIR}}) +1.30_{-0.03}^{+0.03} & (277)\\
\log(\text{SFR}_{\text{UV}}) &= 0.04_{-0.1}^{+0.1}\log(\text{SFR}_{\text{FIR}}) +1.62_{-0.05}^{+0.05} & (65)\\
\log(\text{SFR}_{\text{UV}}) &= 0.2_{-0.2}^{+0.1}\log(\text{SFR}_{3\text{GHz}}) + 1.74_{-0.05}^{+0.04}& (35)\\
\log(\text{SFR}_{\text{MIR}}) &= 0.7_{-0.2}^{+0.2}\log(\text{SFR}_{\text{FIR}}) + 0.5^{+0.4}_{-0.4} & (85)\\
\log(\text{SFR}_{\text{MIR}}) & = 0.8_{-0.1}^{+0.2}\log(\text{SFR}_{3\text{GHz}})+ 0.5^{-0.4}_{+0.3} & (41)\\
\log(\text{SFR}_{\text{FIR}})& = 0.6_{-0.2}^{+0.2}\log(\text{SFR}_{3\text{GHz}})+0.7^{+0.3}_{-0.3} & (41)
\end{align*}

A large scatter exists between the SFR$_{\text{UV}}$ and longer wavelength SFR tracers. The UV suffers from dust attenuation that is dependent on viewing angle amongst other factors. The data in Figure \ref{sfrcomp} are coloured by their inclination angle, and it is clear that the SFR$_{\text{UV}}$ strongly depends on inclination. Face-on galaxies (1-cos($i$)=0) having SFR$_{\text{UV}}$ that lie closer to SFR$_{12\mu\text{m}}$ values in the local sample. However, in almost all galaxies, in both the local and COSMOS samples, SFR$_{\text{UV}} <$ SFR$_{\text{MIR}}$ as there will be some level of FUV attenuation even for face-on galaxies (as discussed in Sects. 3.2 and 4).

In the COSMOS sample at $z\sim0.7$, we find a good agreement between the 24 $\mu$m, 100 $\mu$m, and 3 GHz SFRs. This is encouraging as all three are tied to TIR calibration (through SED fitting or via the IR-radio correlation). 
Locally the SFR$_{\text{FIR}}$ is greater than the SFRs derived from 12 $\mu$m and 1.4 GHz. The 1.4GHz SFR calibration of \cite{Davies2017} used for the local sample depends sub-linearly on SFR. Therefore it follows that 60$\mu$m SFR is higher than the 1.4 GHz SFR because there is a linear relation between 1.4 GHz and FIR (FIRRC). However, this would imply that the FIR emission traced by 60 $\mu$m does not correlate linearly with SFR. On the other hand, our best fits show that SFR$_{12\mu\text{m}}$ is also larger than SFR$_{1.4\text{GHz}}$ at high SFRs. This could imply that the SFR$_{1.4\text{GHz}}$ relation used by \cite{Davies2017} underestimates the SFR.

\section{Evolution of UV attenuation with redshift: Studies included in Figure \ref{evolution}.}\label{studies}

\cite{Bouwens2016} give a ``consensus relationship" for IRX-M$_*$ between $2<z<3$ by combining the results of \cite{Whitaker2014}, \cite{Reddy2010}, and \cite{AlvarezMarquez2016}. In Figure \ref{evolution}, we show the resulting FUV attenuation obtained by evaluating the ``consensus'' equation at  $\log(M_*/M_\odot)$=10.2-10.6.
\cite{McLure2017} fit of the $A_{1600} - M_*$ relation and found the IRX -- $M_*$ relation for star-forming galaxies in the Hubble Ultra Deep field is well described by a \cite{Calzetti2000} attenuation law.
As done for \cite{Bouwens2016}, we show the expected SFR$_{\text{UV}}$/SFR$_{\text{tot}}$ for galaxies with $M_*$ ranging from $10^{10.2}$ to $10^{11.4}$ M$_\odot$ according to the IRX-M$_*$ fit of \cite{McLure2017} as a shaded region.

\cite{Whitaker2014} report the average $L_{\text{IR}}$, $L_{\text{UV}}$ and SFRs for SF galaxies selected via colour-colour criterea in the CANDELs fields in bins of stellar mass and redshift. From these values we calculate SFR$_{\text{UV}}$/SFR$_{\text{tot}}$ and in Figure \ref{evolution} we show the minimum and maximum values found for galaxies with 10.2$<\log(M_*/\mathrm{M_\odot})<$11.4 (to match this work).
For \cite{Wuyts2011b} we read off the SFR$_{\text{UV}}$/SFR$_\text{IR}$ fraction at the position of the MS at log($M_*$/M$_\odot$)=10.2, 10.8, and 11.4 from their Figure 6. Their MS was approximated with a constant slope of 1. We consider the total SFR to be SFR$_{\text{UV}}$+SFR$_{\text{IR}}$ and then solve for SFR$_{\text{UV}}$/SFR$_{\text{tot}}$. The error bars show the range corresponding to the \cite{Wuyts2011b} results at the different stellar masses spanned by this work. 

\cite{Wang2016} give the observed uncorrected FUV luminosity and corrected luminosities for galaxies in their Figure 9 for a sample of galaxies in GAMA/H-ATLAS with slightly lower masses than our sample, median log($M_*$/M$_\odot$) = 10.13, and a median redshift z=0.077. 

\cite{Pannella2009} use BzK selection to perform a stacking analysis on 1.4 GHz data to infer the total SFR at $z\sim2$. Panella report the attenuation at 1500\AA, derived from the stacked SFR$_{\text{FUV}}$/SFR$_{1.4\text{GHz}}$, as a function of stellar mass. In Figure \ref{evolution}, we include the results from 10.2 to 11.4 as indicated by the error-bars. 
\cite{Pannella2015} take a mass-complete sample of galaxies in GOODS-N classified as star forming by a UVJ colour-colour selection. We take measurements of $A_{\text{FUV}}$ at 10.2$<\log(M_*/M_\odot)<$11.4 from Figure 7 in \cite{Pannella2015}. They find that $A_{\text{FUV}}$ increases by 0.3 dex from $z\sim0.7$ to $z\sim1$ and then stays constant out to $z\sim3.3$.

 
\cite{Burgarella2013} and \cite{Cucciati2012} use a ratio of UV and FIR luminosity functions in the VIMOS-VLT Deep Survey to calculate $A_{\text{FUV}}$. This considers galaxies of all stellar masses. The difference between these studies is that \cite{Burgarella2013} uses a PACS-selected sample and \cite{Cucciati2012} use an I-band selected catalogue. 

\cite{Reddy2012} use Herschel measurements of UV-selected star-forming $1.6<z<2.6$ galaxies in GOODS-North. In Figure \ref{evolution}, we show the SFR$_{\text{UV}}$/SFR$_{\text{tot}}$ for their samples A (All UV-selected galaxies) and E (Ultra-Luminous Infrared Galaxies; ULIRGs), where the ULIRG sample has SFR$_{\text{UV}}$/SFR$_{\text{tot}}$ ratios consistent with massive galaxies.
\cite{Heinis2013} calculated $A_{\text{FUV}}$ for a UV-selected sample of galaxies $1.2<z<1.7$ from IR and UV data.
\cite{Dahlen2007} draw on GOODS-South photometry to measure the rest-frame 1500 and 2800\AA~ luminosity functions. The UV slope is used, assuming the Calzetti attenuation law to calculate the FUV attenuation correction factor required. 

\cite{Hao2011} use a number of normal star-forming nearby galaxies from \cite{Kennicutt2003} and \cite{Moustakas2006}. The sample has $\log$(SFR/M$_\odot$yr$^{-1}$) values from -3 to 1. 
\cite{Calzetti2000} combined IR and UV data for a sample of eight nearby galaxies with $\log$(SFR/M$_\odot$yr$^{-1}$) ranging from -1 to 1.7.

\cite{Buat2015} calculate the FUV attenuation of IR-selected galaxies from $z\sim 0.2$  to $z\sim2$ by fitting galaxy SEDs from UV to IR wavelengths. They conclude that galaxies selected in the IR show larger attenuation than galaxies selected in the UV or optical as they are the more massive galaxies (average M$_*>10^{10.4}$ M$_\odot$). In Figure \ref{evolution}, we show the average $A_{\text{FUV}}$ for each redshift bin quoted in Table 3 of \cite{Buat2015}. The error bars show the sample dispersion of 1.3 mags.

\end{document}